%% file: main-1124.tex
\documentclass[aps,prd,twocolumn,nofootinbib,superscriptaddress,amsmath,amssymb,floatfix,10pt]{revtex4-1}

\usepackage{graphicx} % Include figure files
\usepackage{dcolumn} % Align table columns on decimal point
\usepackage{bm}      % bold math
\usepackage{hyperref} % add hypertext capabilities

\usepackage{adjustbox}
\usepackage{multirow}
\usepackage[title]{appendix}

\input{./def.tex}

\newcommand{\GaTech}{\affiliation{School of Physics, Georgia Institute of Technology, Atlanta, Georgia 30332, USA}}

\newcommand{\Yale}{\affiliation{Department of Physics, Yale University,New Haven, CT 06511, USA}}
\newcommand{\CAU}{\affiliation{Department of Physics, Chung-Ang University, Seoul 06974, Republic of Korea}}

\begin{document}

%\title{Detectability of Gravitational-Wave Emission from BBH Hyperbolic Encounters in Ground-Based Interferometers}
\title{A Model-Independent Framework for Gravitational-Wave Reconstruction of Binary Black Hole Hyperbolic Encounters in Ground-Based Interferometers}
\author{Peter Lott}\GaTech\affiliation{Phenikaa Institute for Advanced Study, Phenikaa University, Duong Noi, 12116 Hanoi, Vietnam}
\author{Heleen Amedi}\GaTech
\author{Jay Graves}\Yale
\author{Yeong-Bok Bae}\CAU
\author{Margaret Millhouse}\GaTech
\author{Laura Cadonati}\GaTech

\date{\today}

\begin{abstract}
Binary black hole hyperbolic encounters represent a dynamical interaction in which two black holes undergo a close fly-by, emitting gravitational-wave bremsstrahlung in the form of a short-duration, single-cycle transient. These events are expected to occur in dense stellar environments such as globular clusters and both active and quiescent galactic nuclei. 
In this work, we constrain the detection sensitivity for hyperbolic encounters of black hole pairs with a range of asymmetric masses.
We employ \bw{}, a wavelet-based morphology-independent algorithm to characterize hyperbolic encounter waveforms in simulated detector noise; for this study, we explore the use of exponential shapelets.
We find that a typical hyperbolic orbit with total mass $20\,\msol$ can be detected up to distance $d_L \sim 40 - 200$ Mpc, and we forecast the possibility of detection by ground-based current and future gravitational wave interferometers. 
\end{abstract}

\maketitle

%%%%%%%%%%%%%%%%%%%%%%%%%%%%%%%%%%%%%%%%%%%
\section{Introduction}
%%%%%%%%%%%%%%%%%%%%%%%%%%%%%%%%%%%%%%%%%%%
The \ligo{}, Virgo and KAGRA detectors (LVK) have to date published more than 200 \gw{} events from mergers of compact objects in bound orbits~\cite{LIGOScientific:2021djp, LIGOScientific:2025slb}. 
Most of these events are \bbh{} coalescences, although systems with neutron stars have also been detected~\cite{LIGOScientific:2021qlt,LIGOScientific:2017vwq, LIGOScientific:2020aai,LIGOScientific:2020zkf}.
The increasing sensitivity of current ground-based detectors, as well as prospects for next-generation interferometers such as Cosmic Explorer (CE)~\cite{Reitze:2019iox} and Einstein Telescope (ET)~\cite{Maggiore:2019uih}, have widened the potential discovery window. 
For example, recent work proposed \bbh{} hyperbolic encounters as a viable source of GW radiation detectable in ground-based interferometers~\cite{10.1093/mnras/stab2721,Caldarola:2023ipo, Morras:2021atg,Bini:2023gaj}. In two-body fly-by interactions,
\gw{s} are emitted as
\textit{\gw{} bremsstrahlung}; 
for binary masses in the \ligo{} band~\cite{Capozziello:2008mn} if the encounters are sufficiently close, this radiation can reach the sensitivity band of ground-based interferometers like LIGO and Virgo.
As a \gw{} source, \bbh{} hyperbolic encounters are a unique probe for black holes, as their extreme eccentricity can constrain formation channels, in conjunction with radiation-driven dynamical captures~\cite{Ebersold:2022zvz}.
\bbh{} hyperbolic encounters may also make up a non-negligible portion of the stochastic \gw{} background, particularly in the primordial black hole hypothesis~\cite{Garcia-Bellido:2021jlq, Garcia-Bellido:2017qal, Garcia-Bellido:2017knh}.

Physically, the hyperbolic encounter can be described as a classical two-body problem in an unbound orbit, characterized by the eccentricity $e  \gtrsim 1$, the initial relative velocity $v_{0}$, the total mass $M$ and the impact parameter $b$, a length-scale property of the orbit. As the system evolves, the trajectory of the secondary black hole is perturbed by the gravitational field of the primary black hole, resulting in a {\em braking radiation} analogous to electromagnetic bremsstrahlung. Thus, the problem can be described with a relatively small number of parameters whose waveforms may be approximated by similar methods as in the merger case~\cite{Capozziello:2008mn}. For this reason, the hyperbolic two-body problem has seen treatment in post-Newtonian~\cite{Hansen1972, 1977ApJ...216..610T, Cho:2018upo, Cho:2022syn} and post-Minkowskian~\cite{1975ApJ...200..245T, 1977ApJ...215..624C, 1977ApJ...217..252K, 1978ApJ...224...62K, Vines:2017hyw} approximations.
Numerical relativity simulations of \bbh{} hyperbolic encounters in unequal masses~\cite{Bae:2017crk, Bae:2020hla, Bae:2023sww} have revealed a morphology rich in spin effects~\cite{Bae:2020hla}, including ringdown effects~\cite{Bae:2023sww}.

From a data analysis standpoint, \bbh{} hyperbolic encounter \gw{} signals are morphologically distinct from their merger counterparts: they are single-cycle, i.e., broadband, as opposed to quasiperiodic \bbh{} merger waveforms. In addition, their waveforms exhibit the linear memory effect, which binary mergers typically do not due to their rotational symmetry, although there are exceptions~\cite{Favata:2011qi, PhysRevD.45.520}. 
A search has been carried out for \bbh{} hyperbolic encounters in data from the third LVC observing run (O3)~\cite{Morras:2021atg, Bini:2023gaj}, but no significant events have been confirmed. The rate estimates show that they can reach a few per year~\cite{10.1093/mnras/stab2721, Kocsis:2006hq, Capozziello:2008mn} in the upgraded version of LIGO, A+~\cite{barsotti2018a+}.  

In this work, we present a study of simulated signals of \gw{} radiation from black holes in unbound orbits. The motivation for this study is two-fold. For one, the inner dynamics of globular clusters and active galactic nuclei disk scenarios could be a perfect environment for fly-by orbits between compact objects, as has been demonstrated in several many-body simulations~\cite{Quinlan1989, Li2022, Codazzo:2022aqj}. Detection of \gw{} emission could provide insight into black hole populations in these environments~\cite{10.1093/mnras/stab2721}. For another, \bbh{} hyperbolic encounters present a challenge in data analysis. Given that they are morphologically distinct from the chirp-like profile exhibited by circular binaries, \bbh{} hyperbolic encounters may provide a unique opportunity to test the waveform systematics of model-dependent and independent algorithms alike.

\begin{figure}[th]
%\hspace*{-1.3cm} 
    \includegraphics[width=0.45\textwidth]{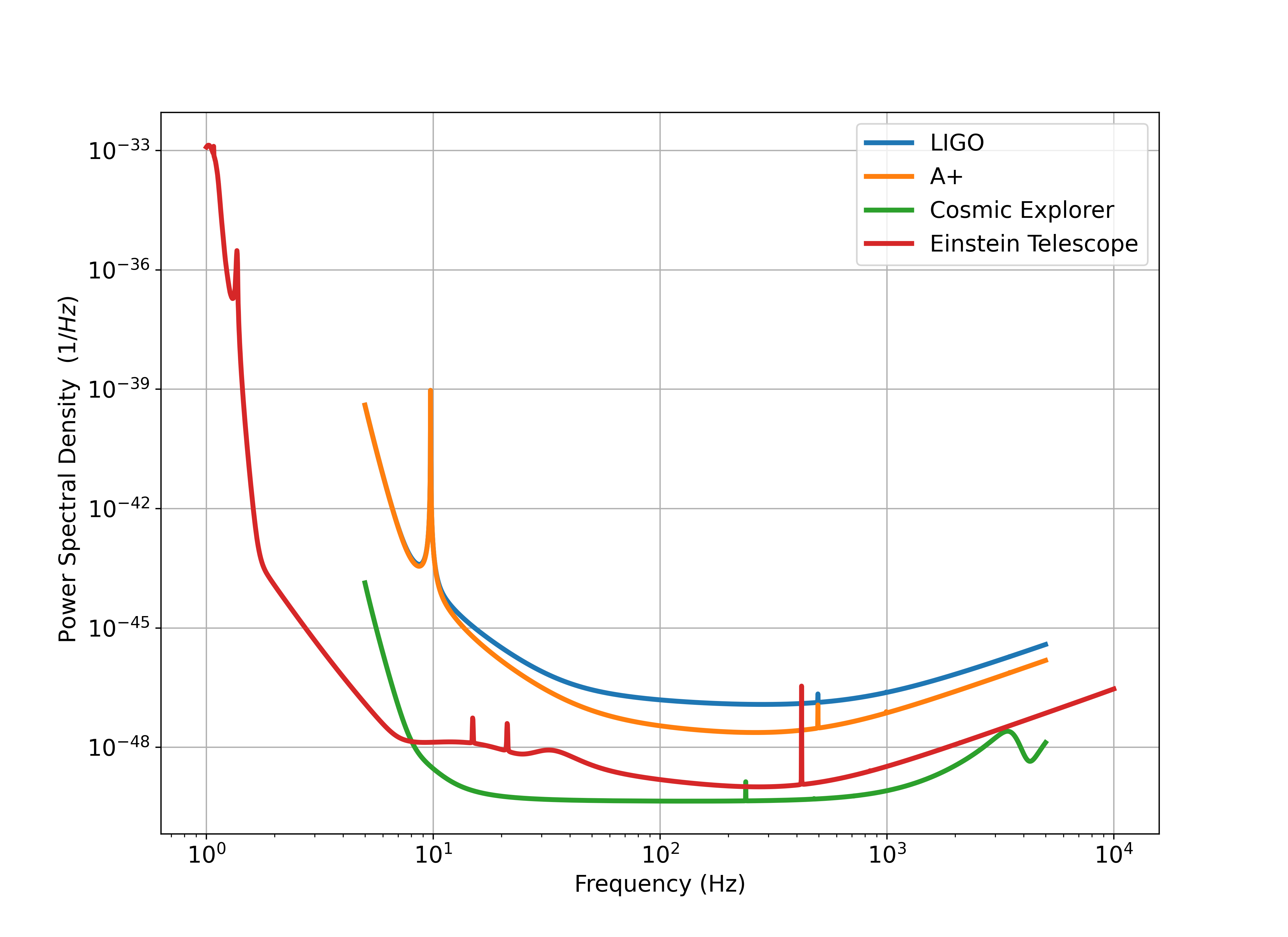}
    \caption{Sensitivity curves for LIGO~\cite{Cahillane:2022pqm}, A+~\cite{barsotti2018a+}, Cosmic Explorer~\cite{Reitze:2019iox}, and Einstein Telescope~\cite{Maggiore:2019uih} (data from\cite{T1500293}).  
    Previous estimates~\cite{Kocsis:2006hq, Bini:2023gaj} suggest that hyperbolic encounters are most common around $10$Hz. Detectors with improved low-frequency sensitivity, such as Cosmic Explorer and the Einstein Telescope, are better suited to capturing these signals than current-generation instruments like LIGO or even A+.
  }
    \label{fig:PSDs}
\end{figure}

In this paper, we employ waveforms from a numerical simulation of \bbh{} hyperbolic encounters with unequal masses~\cite{Bae:2017crk}. The waveforms are quadrupolar-mode and spinless, spanning a grid of waveform parameters. To reconstruct the waveform we use \bw{}~\cite{Cornish:2014kda, Cornish:2020dwh}, a model-agnostic Bayesian inference algorithm that resolves features in the data without assuming a-priori knowledge of the signal. We use these simulated signals (hereafter referred to as \textit{injections}) to empirically constrain detection rates for LIGO, A+ CE and ET~\cite{Moore:2014lga}. 
 The strain sensitivity for the four detectors is presented in Figure \ref{fig:PSDs}. 

For this exploratory work, we employ \textit{exponential shapelets}~\cite{Berge:2019nyt} as  frame functions and compare their reconstruction faithfulness to frames previously used in morphology-independent analyses~\cite{Cornish:2014kda, 2016PhRvD..93b2002K, Drago:2020kic,  Bini:2023gaj, Ghonge:2020suv}.

%%%%%%%%%%%%%%%%%%%%%%%%%%%%%%%%%%%%%%%%%%%
\section{GW Burst Reconstruction}
%%%%%%%%%%%%%%%%%%%%%%%%%%%%%%%%%%%%%%%%%%%

\subsection{\bw{}}

\bw{} is a model-agnostic algorithm for gravitational-wave inference that reconstructs transient features in detector data using a sum of frame functions~\cite{Cornish:2014kda, Millhouse:2018dgi, Cornish:2020dwh}. 
Unlike templated analyses~\cite{Biwer:2018osg, PhysRevD.91.042003, 2021SoftX..1400680C},
%\LC{citation needed}\PL{Added RIFT, GstLAL, LALinference and pycbc inference citations}
BayesWave does not return posterior distributions of the physical parameters of the gravitational-wave signal; rather, it returns a posterior distribution of the signal $\mathbf{h}(t)$ itself, via the Bayes' Theorem:
\begin{equation}
    p(\mathbf{h} | \mathbf{d}) = \frac{p(\mathbf{h}) p(\mathbf{d} | \mathbf{h})}{p(\mathbf{d})},
\end{equation}
where $\mathbf{d}$ is the detector data (which, in the absence of a signal, we assume to be stationary and Gaussian), $p(\mathbf{h})$ is the prior knowledge about the system, $p(\mathbf{d} |\mathbf{h})$ is the likelihood, 
and $p(\mathbf{d})$ is the evidence. 
BayesWave draws from this posterior $p(\mathbf{h}|\mathbf{d})$  using a reversible-jump Markov chain Monte Carlo (RJMCMC), which samples over the frame function parameters as well as the number of frame functions. By using an RJMCMC that allows us to sample transdimensionally (i.e., adding or subtracting frame functions), \bw{} can faithfully reconstruct a wide range of signal morphologies, may fair better than template-based reconstructions in situations where information about the model is incomplete
(e.g. higher order modes or physics beyond GR)~\cite{Cornish:2014kda, Millhouse:2018dgi, Ghonge:2020suv, Cornish:2020dwh}.
Due to its inherent flexibility, \bw{} has been used in the processing of events in \gw{} transient catalogs~\cite{LIGOScientific:2018mvr, LIGOScientific:2020ibl, LIGOScientific:2021usb, KAGRA:2021vkt}, for tests of general relativity~\cite{LIGOScientific:2019fpa, LIGOScientific:2020tif}, and in the mitigation of instrumental glitches~\cite{PhysRevD.98.084016, Chatziioannou:2021ezd, Ghonge:2023ksb}.

Two frame function classes have been previously implemented in \bw{}: sine-gaussians (also called {\it Morlet-Gabor} wavelets), and {\it chirplets}~\cite{Cornish:2020dwh, Millhouse:2018dgi}. 

Morlet-Gabor wavelets are maximally compact in time-frequency, and have an analytic expression for the Fourier transformation, which is beneficial since \bw{} calculates the likelihood in the frequency domain~\cite{Cornish:2014kda}. Their time-domain expression is: 
\begin{flalign}
& \Psi_{\rm Morlet-Gabor}(t; t_0, f_0, Q, A, \phi_0) = \nonumber \\ 
& \hspace{2cm} Ae^{-(t - t_0)^2/\tau^2} \cos({2 \pi f_0 (t - t_0) + \phi_{0}})
\end{flalign}
\noindent
where $t_0$ and $f_0$ are the central time and the central frequency, $\phi_0$ is the phase offset, $A$ is the amplitude, and $Q = \Delta t \Delta f$, is the \textit{quality factor}.
We also define $\tau = Q/2 \pi f_0$ for convenience.
Their Fourier transform is:
\begin{flalign}
& \Psi_{\rm Morlet-Gabor}(f; t_0, f_0, Q, A, \phi_0) = \nonumber \\ 
&  \frac{\sqrt{\pi} A \tau}{2}  
  e^{-\pi^2 \tau^2 (f - f_0)^2} e^{-2 \pi i f t_0} (e^{i \phi_0} + e^{-i \phi_0} e^{- Q^2 f / f_0}).
\end{flalign}

Chirplets  mathematically differ from Morlet-Gabor wavelets in the time domain by a frequency evolution term which allows for the chirplet's frequency to smoothly increase or decrease over time~\cite{Millhouse:2018dgi, Chassande-Mottin:2010hsa}:
%$\Psi_{\rm chirp}(f;t_0,f_0,Q,A,\phi_0,\dot{f_0})$: 

%
\begin{flalign}
& \Psi_{\rm chirplet}(t; t_0, f_0, Q, A, \phi_0, \dot{f_0}) = \nonumber \\ 
&  \frac{\sqrt{\pi} A \tau}{2}  
  Ae^{-(t - t_0)^2/\tau^2} \cos({2 \pi f_0 (t - t_0) + \pi \dot{f_0} (t - t_0)^2\phi_{0}}),
  \label{eq:chirplets}
\end{flalign}
where $\tau = Q / 2 \pi f_0 $ The parameter $\dot{f_0}$, can be positive or negative and describes the linear increase or decrease of frequency over time. 
The analytical expression for chirplets in the frequency domain can be found in~\cite{Millhouse:2018dgi}.

Morlet-Gabor wavelets and chirplets have both been shown to accurately reconstruct simulated GW signals~\cite{Becsy2017,Pannarale2018,Ghonge:2020suv,Becsy2020}. For compact binary coalescences, which increase in frequency up to the merger, the added flexibility of the $\dot{f_0}$ parameter in chirplets leads to larger overlaps between the injected signal and the waveform recovered by BayesWave.  However, for white noise burst test signals, with no well defined frequency evolution, chirplets and wavelets have been shown to perform comparably~\cite{Millhouse:2018dgi}.   
Outside of BayesWave,  Morlet-Gabor and chirplet bases have proven beneficial in time-frequency \gw{} \cbc{} analysis, particularly with modulation $Q$ to extract the postmerger component of a NS-NS binary~\cite{2024arXiv240216533H}.

%\sout{It is computed according to the standard calculations which take into account thermal noise, and quantum shot noise} 
%\MM{I still think this last sentence about ``computing'' and ``standard calculations'' of the PSD is confusing.  The thermal noise, shot noise, etc are used to \emph{model} what the PSD should look like, but in actual data analysis the PSD is estimated from the data itself.}
%\MM{projected PSDs are calculated using thermal, quantum, etc noise. For real data, the PSD is just estimated off the data itself.}. 

%\MM{Probably don't need Eq 7}In the case of SNR that reduces to 
%\sout{\begin{equation}
%    \mathrm{SNR}^2 = 4\int_{0}^{\infty}\frac{\tilde{h}^* \tilde{h}}{S_n(f)}df.
%\end{equation}}

%\bw{} approximates the noise curve by taking a Gaussian distribution about that curve and uses that to calculate the PSD and SNR of the signal. 
%\LC{this is not very clear}

\subsection{Shapelets}

 In this work, we investigate  {\it shapelets}~\cite{Berge:2019nyt, Baghi:2021tfd} as a new set of \bw{} frame functions, following their prior use in searches for gravitational lensing~\cite{Refregier:2001fd, Refregier:2001fe, 2015ApJ...813..102B}, exoplanets~\cite{2005ApJ...626.1070H, 2012MNRAS.427..948A}, pulsar timing~\cite{2016PhRvD..93h4048E, 2016MNRAS.458.3341D}, and other astrophysical phenomena~\cite{2005MNRAS.363..197M}. 
 Shapelets are Hermite functions multiplied by an envelope function.
 We employ the \textit{exponential shapelets} variant~\cite{Berge:2019nyt, Baghi:2021tfd} recently used in a LISA Pathfinder glitch analysis~\cite{Baghi:2021tfd}. Using an exponential envelope introduces heavier tails and asymmetry, making exponential shapelets better suited  for capturing sharp or extended \gw{} features~\cite{Berge:2019nyt}, compared to the more localized Gaussian shapelets~\cite{Cornish:2014kda, Millhouse:2018dgi, Drago:2020kic}.

Shapelets are a complete and orthogonal set~\cite{Refregier:2001fd, Berge:2019nyt}. 
Their time-domain expression is: 
\begin{flalign}
&  \Psi_{shape}(t; t_0, f_0, Q, A) = \nonumber \\ 
&  \frac{(-1)^{n - 1}A}{\sqrt{n^3 \beta}} \left( \frac{2 (t-t_0)}{n \beta}\right)  
   L^{1}_{n - 1}  \frac{2 (t - t_0)}{n \beta} e^{-(t-t_0) / {n \beta}} e^{-2\pi i t f_0},
   \label{eq:shapeletsTimeDomain}
\end{flalign} 

\noindent
where $\beta = Q/{2 \pi f_0}$, $L_{n - 1}^{1}(t-t_0) $ is the first-order generalized Laguerre polynomial.
%and \LC{but you only use the first order polinomial, so why introducing $\alpha$?} \PL{I corrected this to only refer to the first polynomial.}
In the frequency domain, they can be expressed as:

\begin{flalign}
&  \tilde{\Psi}_{shape}(f; t_0, f_0, Q, A) = \nonumber \\ 
&  (-1)^n A \sqrt{\frac{2n\beta}{\pi}}  
   \frac{\left(n\beta(f-f_0)-i\right)^{2n}}{\left((n\beta(f-f_0)^2+1)\right)^{n+1}}e^{-2\pi i t_0 f}.
\end{flalign}

\noindent
%\MM{ Eq 5 is only valid for $t>t_0$. There should be some discussion about this and how it's handled.} 
%\PL{The analysis is largely in the frequency domain using $\tilde{\Psi}_{shape}(f)$, and the inverse Fourier transform is applied at the end solely for visualization as a post-processing step. The phase factor $\exp(-2\pi i f t_0)$ in the frequency domain time-shifts shifts waveform, ensuring that the time-domain signal is strictly zero for $t < t_0$. So the frequency-domain formulation accurately represents the intended temporal behavior without introducing any unintended signals prior to t0. }
Though Eq.~\ref{eq:shapeletsTimeDomain} is only valid for $t>t_0$, because the analysis is performed in the frequency domain and only inverse Fourier transformed into the time domain for visualization purposes; in practice $\Psi_{\rm shapelet}(t) = 0$ for $t<t_0$.

\begin{figure}[tb]
\includegraphics[width=.45\textwidth]{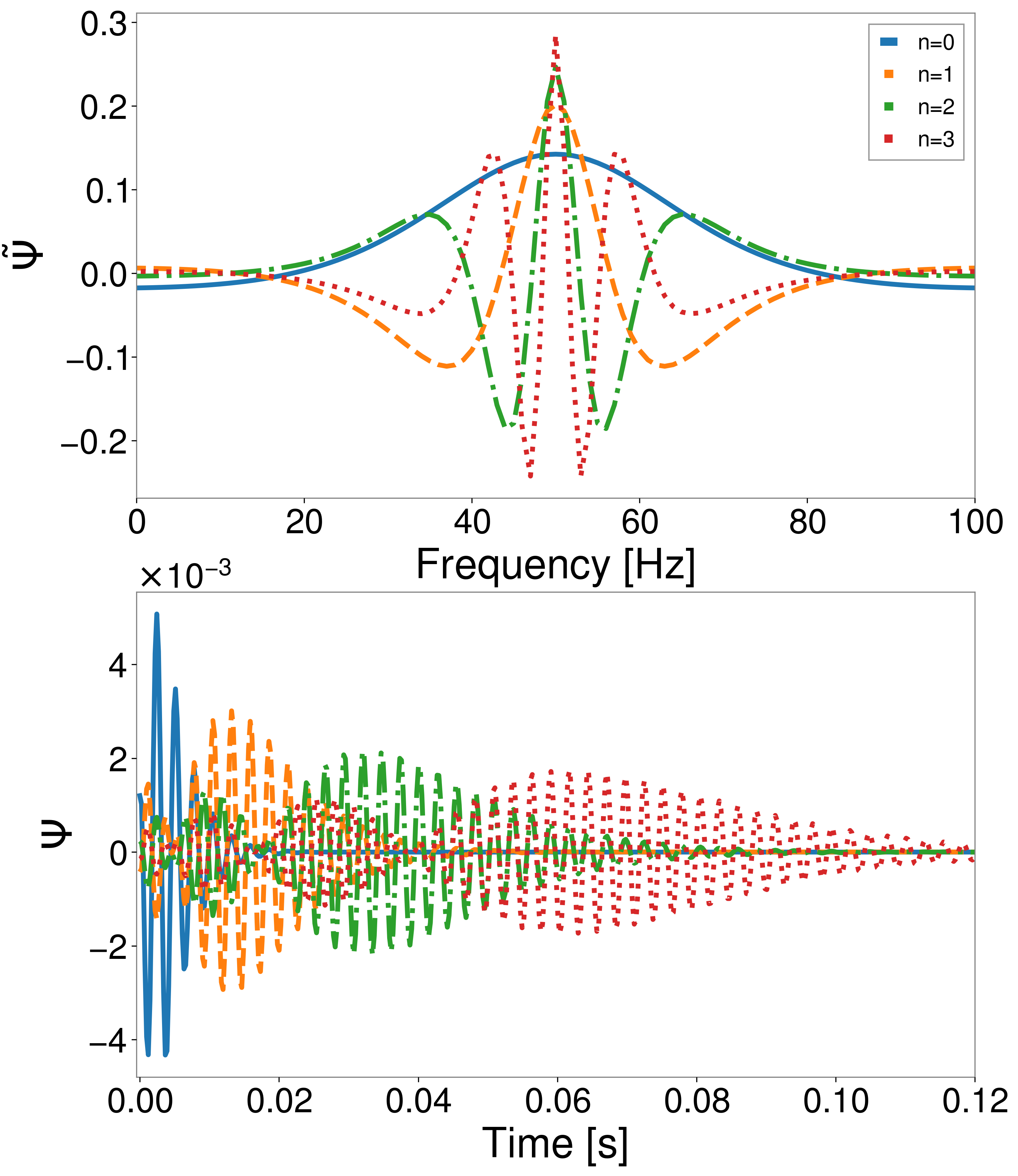}
\caption{Examples of $n=0,1,2,3$ shapelets in the frequency domain (top), and time domain (bottom). For all shapelets, $Q=20$, $f_0=200\mathrm{Hz}$, and $t_0=0\mathrm{s}$.
}
\label{fig:shape_plots}
\end{figure}

For reference, $n=0$  to $n=3$ shapelets $\tilde{\Psi}_{n}$ are plotted in the Fourier and time domains in Figure \ref{fig:shape_plots}.  
As the index number $n$ increases, the Laguerre term in $\tilde{\Psi}_{n}$ dominates and the number of oscillations increases as a function of frequency. 
For simplicity,  this study is restricted to shapelets with $n=2$. 

%In this study, our objective was to accurately capture the complexity of the simulated signal in a manner remains theoretically tractable and interpretable. We focus on $n=2$ because it represents the lowest-order nontrivial case that captures essential interactions. \MM{I'm still a bit confused by why only using n=2. I don't think increasing n should increase the computational costs very much?}. \PL{The n=2 waveform model provides the minimal complexity necessary to accurately capture the distinctive features of BBH hyperbolic encounter signals, which exhibit sharp bursts with asymmetric amplitude and phase evolution and even ringdown effects. The simpler n=1 is basically a gamma distribution function (linear times a decaying exponential) which lacks the flexibility to model these characteristics, resulting in representation that cannot fully reproduce the physical waveform morphology. By incorporating the second-order Laguerre polynomial, the n=2  model captures inflection points for the purpose of more precise reconstruction and interpretation.} Higher orders of $n$ are left for future work as they may complicate interpretation.

The primary motivation for using shapelets is their morphological resemblance to gravitational-wave signals from hyperbolic encounters. Transdimensional MCMCs balance the  complexity of a model (i.e. the number of frame functions used to reconstruct a signal) against how well it fits  the data. Chirplets were motivated by compact binary coalescences, the most abundant \gw{} source, which exhibit continuously increasing frequency, and   indeed they improved  \bw{} reconstructions of binary black hole simulations~\cite{Millhouse:2018dgi}.
 Given the resemblance between shapelets and hyperbolic encounter waveforms, we test whether shapelets outperform wavelets and chirplets at reconstructing simulated hyperbolic signals.

The implementation of shapelets in \bw{} is nearly identical to that of chirplets and Morlet-Gabor wavelets, sampling parameters $t_0, f_0, Q$ and $A$ but without a phase parameter $\phi_0$. We use the same priors  as~\cite{Cornish:2020dwh}: 
flat priors for $t_0$, $f_0$, $Q$, while the prior on $A$ is actually a prior on the frame function's signal-to-noise ratio (SNR). 
The SNR of frame function $\psi$ is  $\mathrm{SNR}=\sqrt{(\psi|\psi)}$, where the round brackets denote the noise-weighted inner product:
\begin{equation} \label{eq:inner_prod}
    (a|b) = 2\int_{0}^{\infty}\frac{\tilde{a}^* \tilde{b}+\tilde{a} \tilde{b}^*}{S_n(f)} df,
\end{equation}
where $S_n(f)$ is the one-sided noise power spectral density.
For shapelets, we use the approximation:
\begin{equation}
    \mathrm{SNR}^2 \approx \frac{4A^2}{S_n(f_0)}.
\end{equation}
For all wavelet models, the prior on the number of wavelets $D$ is empirically derived from BayesWave’s unmodeled search of data from~\ligo{'s} first observing run:

\begin{equation}
    p(D) = \frac{4 \sqrt{3} D}{2 \pi a^2 (3 + \frac{D}{a})^4},
    \label{eq:wavletPrior}
\end{equation}
with $a = 2.9$~\cite{Cornish:2020dwh}.

\begin{figure}[t]
    \centering
    \includegraphics[width=0.45\textwidth]{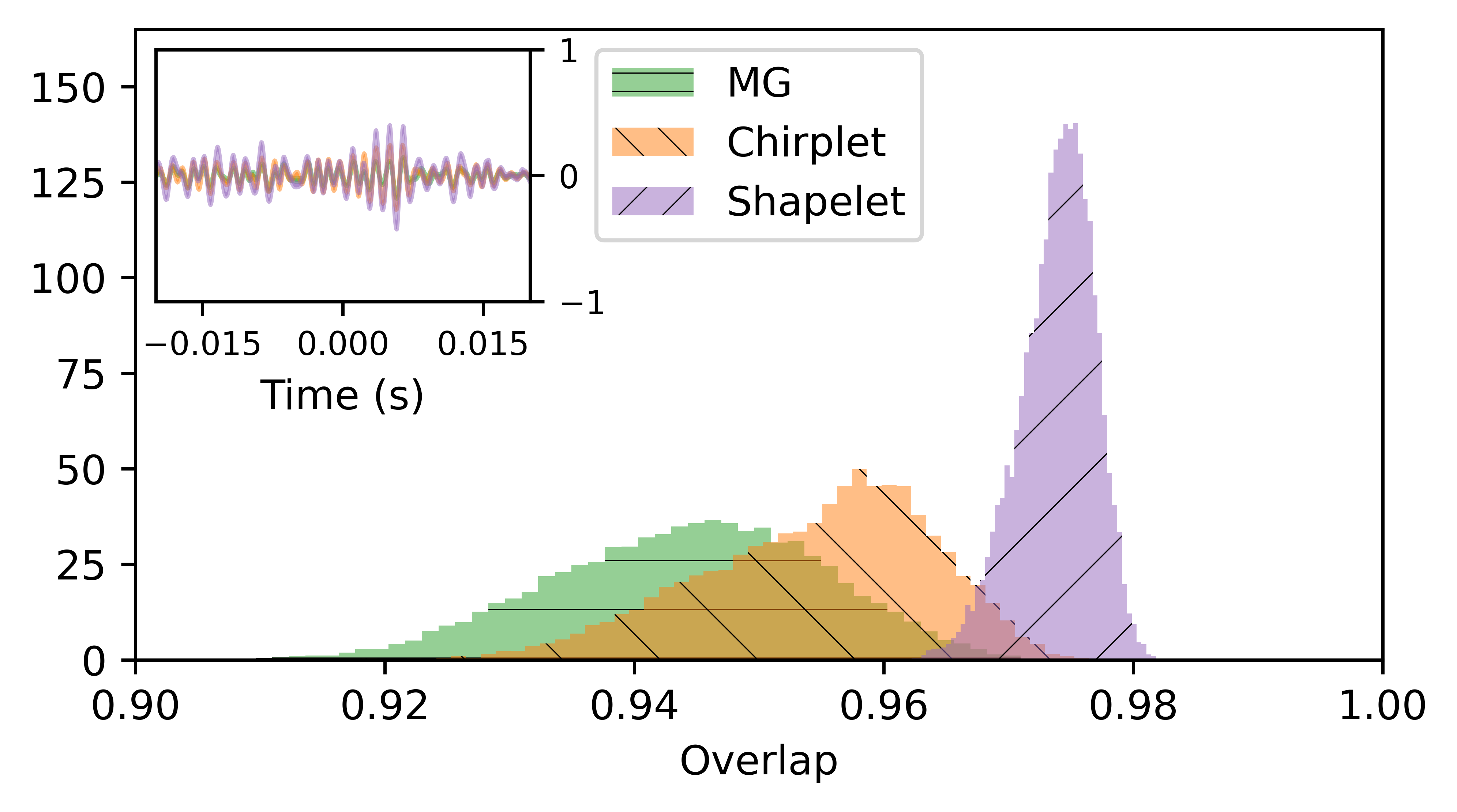}
    
    \caption{Example \bw{} waveform reconstruction of simulated white noise bursts (WNBs) with the three frame function families used in this study. WNBs are unpolarized signals with complicated frequency structure, (due to their localization in time) providing a a good test for the faithfulness of \bw{'s} reconstruction. 
    }
    \label{fig:WNB}
\end{figure}

\begin{figure}[bt]
\hspace{-1.5cm}
        \includegraphics[width=0.45\textwidth]{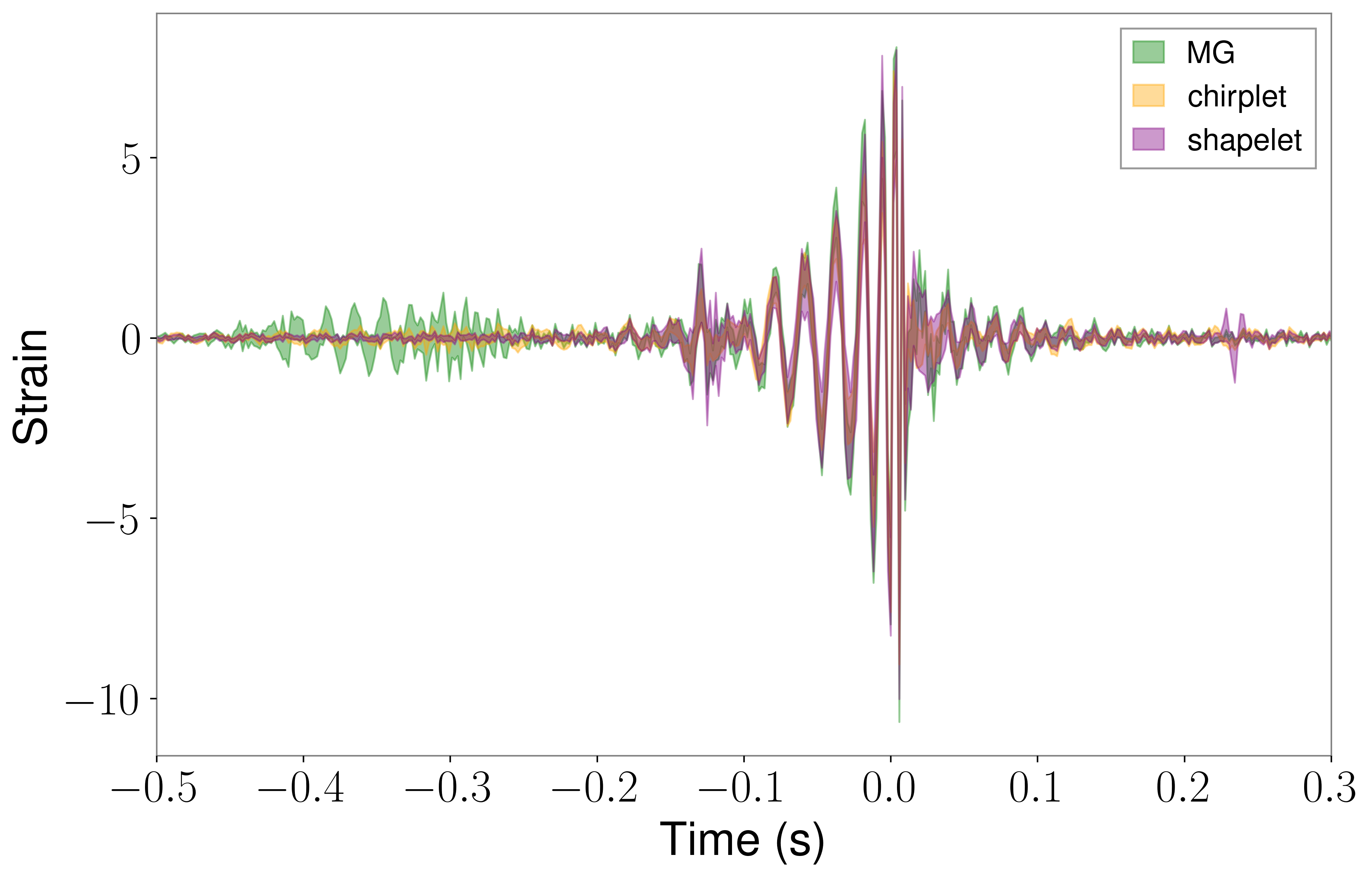}
        \caption{Time-domain reconstruction of GW150914, the first \gw{} event, using \bw{} with shapelets and the standard Morlet-Gabor wavelet reconstructions, respectively. This analysis corresponds to LIGO-Hanford.}
        \label{fig:GWsignal}
    \end{figure}

\subsection{Comparing frame functions} 

We compare the performance of shapelets with other \bw{} frame functions using simulated signals and gravitational wave data. 

For simulated signals, we quantify the \bw{} reconstruction accuracy as the \textit{overlap}  between injected ($h_{inj}$) and  reconstructed  ($h_{rec}$) waveforms:
\begin{equation}
{\cal{O}} \equiv \frac{\left( h_{inj} | h_{rec} \right)}{\sqrt{\left( h_{inj} | h_{inj} \right) \left(h_{rec}|h_{rec} \right)}}.
\end{equation}
%
%%%%%%%%% As a test case, we reconstruct \textit{white noise bursts} (WNBs), which are instances of excess Gaussian strain with a limited duration and bandwidth. Fig.~\ref{fig:WNB} shows overlap histograms for the reconstruction of one such simulation where the overlaps are drawn from the posterior distribution produced by \bw{}. We assume simulation parameters central frequency of $150$ Hz, bandwidth $100$ Hz and duration of $0.1$ s.  For this Hanford injection, \LC{does it matter it is Hanford if this is just a simulation?} we assume an idealized noise curve corresponding the LIGO (see \ref{fig:PSDs}) oriented at an optimal sky location and SNR of 30.
As a test case, we reconstruct \textit{white noise bursts}, instances of excess Gaussian strain with limited duration and bandwidth. Fig.~\ref{fig:WNB} shows overlap histograms for reconstructing one such simulation, with   150~Hz central frequency, 100~Hz bandwidth, 0.1~s duration, and SNR=30 in ideal LIGO noise (see Fig.~\ref{fig:PSDs}) with optimal sky location. The overlaps are drawn from the \bw{} posterior distribution. We found similar trends for different parameter choices.  
As they are not tied to any astrophysical model, white noise bursts are well suited to test the robustness and flexibility of model-independent data analysis methods~\cite{Abadie_2012}. 
The overlap distribution for shapelets is centered around larger overlap values than the other two frame functions, suggesting better performance in capturing the structure of this particular simulated white noise burst signal.

%%%%%%%%%%%%%%% Fig.~\ref{fig:GWsignal} shows reconstructed waveforms for GW150914~\cite{PhysRevLett.116.061102}, the first \bbh{} \gw{} event. Its high network SNR of 24 makes it an ideal test case for all three frame functions. Unlike white noise bursts, which have simpler frequency structures, BBH waveforms exhibit smoothly increasing frequencies and more complex patterns. The reconstructed waveforms for Morlet-Gabor, chirplet, and shapelet frames, shown in Fig. \ref{fig:GWsignal}, appear qualitatively similar. To quantify reconstruction quality, we computed the overlaps with the best-fit GW150914 waveform for each frame function, obtaining values of 0.934, 0.935, and 0.928, respectively, with respect to the best-fit waveform, demonstrating comparable performance. We note also that the Morlet-Gabor and shapelet reconstructions show premerger and post-merger artifacts in the waveform, which is common in this type of analysis~\cite{torrence1998practical, mallat1999wavelet}.
%
Fig.~\ref{fig:GWsignal} instead shows the reconstructed waveforms for GW150914, the first discovered \bbh{} \gw{} event detected with a network signal-to-noise ratio of 24~\cite{PhysRevLett.116.061102}. 
Reconstructions using Morlet-Gabor, chirplet, and shapelet frames yield similar results, with overlaps of 0.934, 0.935, and 0.928, respectively, when compared to the best-fit waveform from general-relativity based estimation. We note also that the Morlet-Gabor and shapelet reconstructions exhibit pre- and post-merger artifacts, which is a common feature in wavelet-based analyses~\cite{torrence1998practical, mallat1999wavelet}.
%\MM{added details to what the `best-fit' waveform means because I think that's info the reader might need}
%
These artifacts arise from the overcomplete nature of the frames and the tendency of the model to fit excess power near the signal boundaries, often due to limited time-frequency localization or mismatch between the frame and localized signal features.
It is important to note that the overlaps reported for GW150914 are maximum a posteriori ~\ref{fig:WNB} are derived from the full posterior distribution. The lower overlaps for GW150914 are expected, as it is a real astrophysical signal with complex morphology and sky projection effects, in contrast to the short, high-SNR, and morphologically simpler white-noise burst used in Fig.~\ref{fig:WNB}.

 \begin{figure*}[t!]
        \centering
        \includegraphics[width=0.9\textwidth]{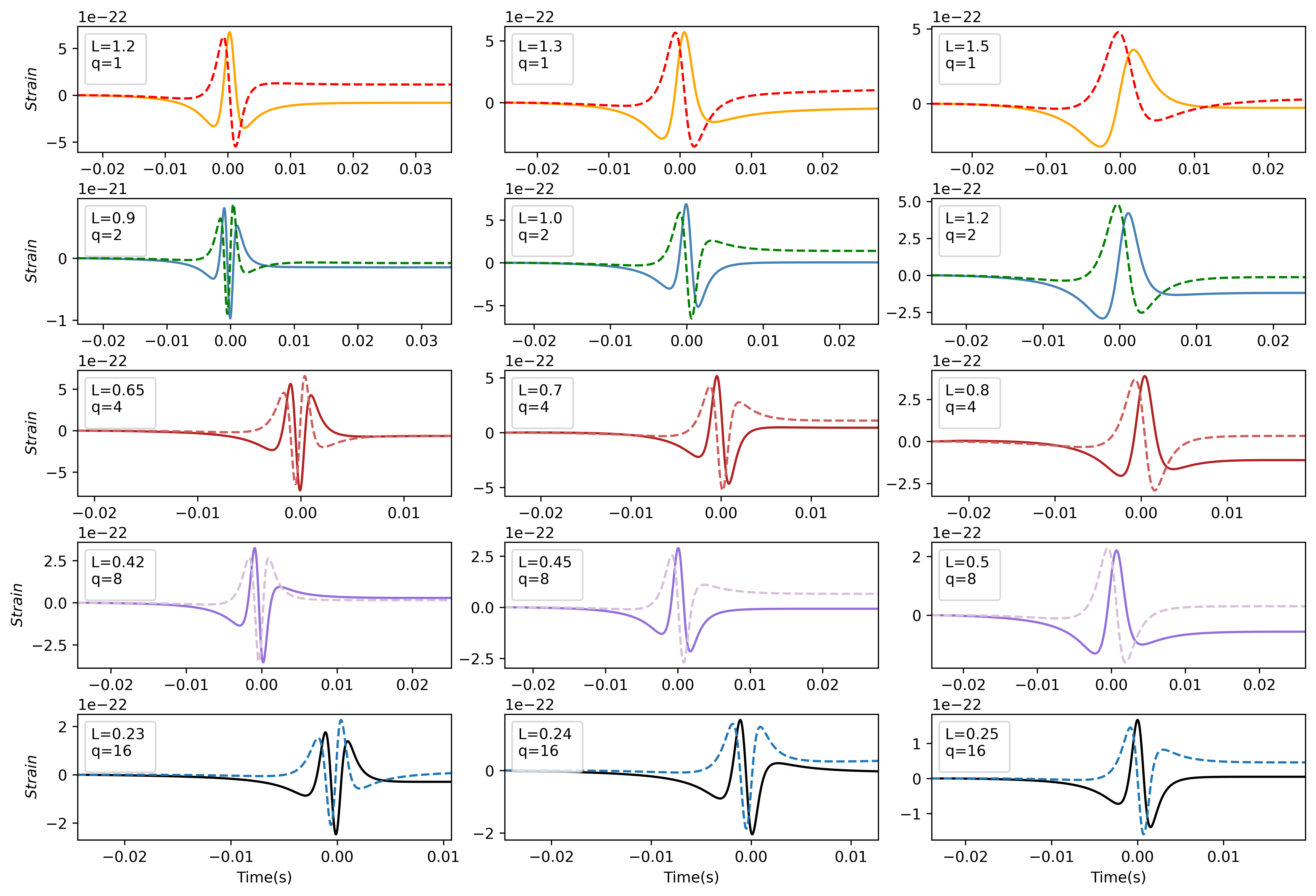}
        \caption{Waveforms used for this analysis, from face-on binaries with total mass $20\,\msol$. Each panel displays is a waveform with plus (solid) and cross (dashed) polarization. This grid spans a parameter space with mass ratios $q = [1, 2, 4, 8, 16]$ (displayed row-wise) and several initial angular momenta, represented column-wise. The waveforms are more narrowly spaced out in time for a larger mass ratio. 
}
        \label{fig:waveforms}
\end{figure*}

%%%%%%%%%%%%%%%%%%%%%%%%%%%%%%%%%%%%%%%%%%%
\section{Hyperbolic Encounters}
%%%%%%%%%%%%%%%%%%%%%%%%%%%%%%%%%%%%%%%%%%%
\subsection{Waveforms}

\begin{figure}[t!]
\includegraphics[width=.45\textwidth]{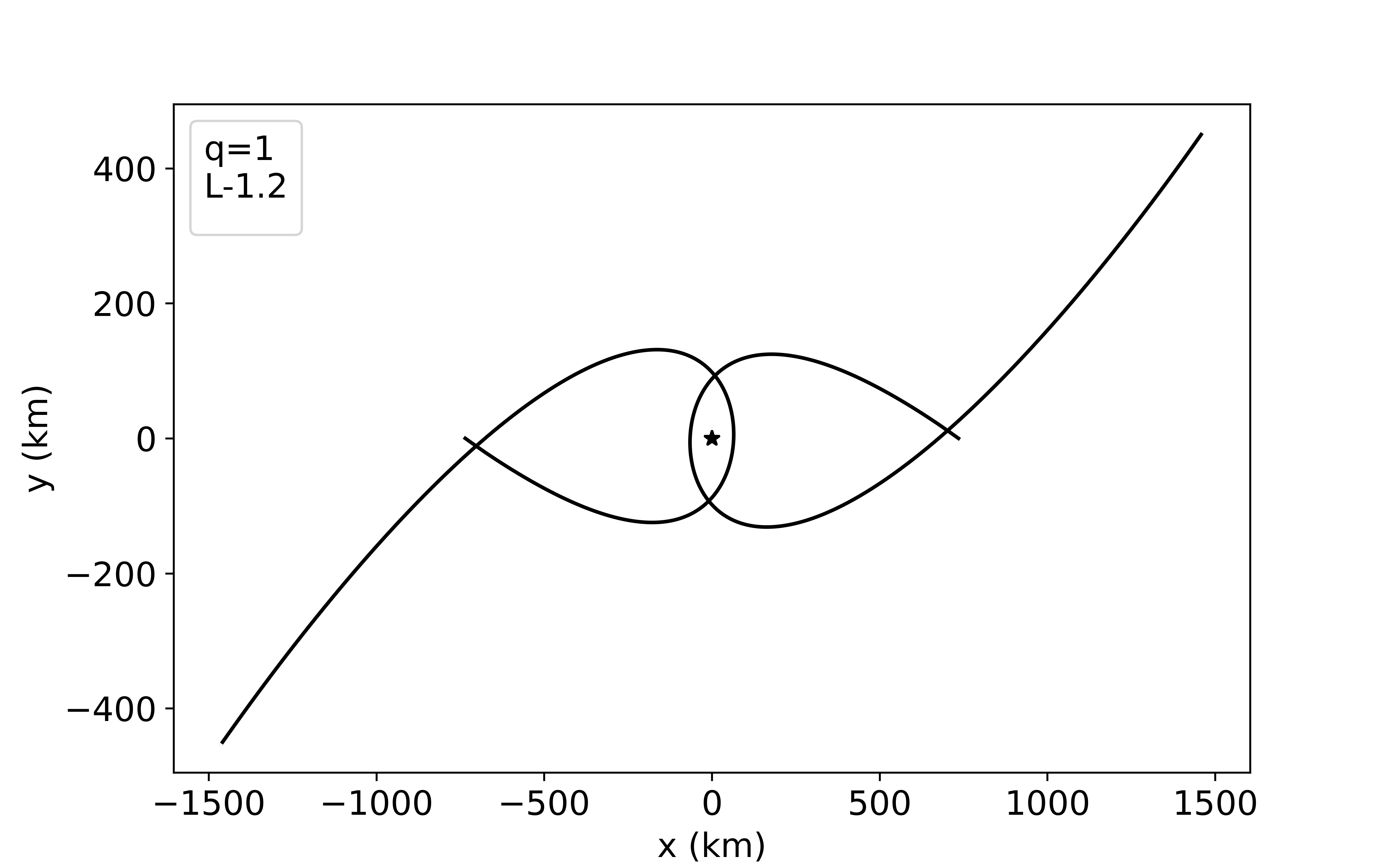}\hfill
\includegraphics[width=.45\textwidth]{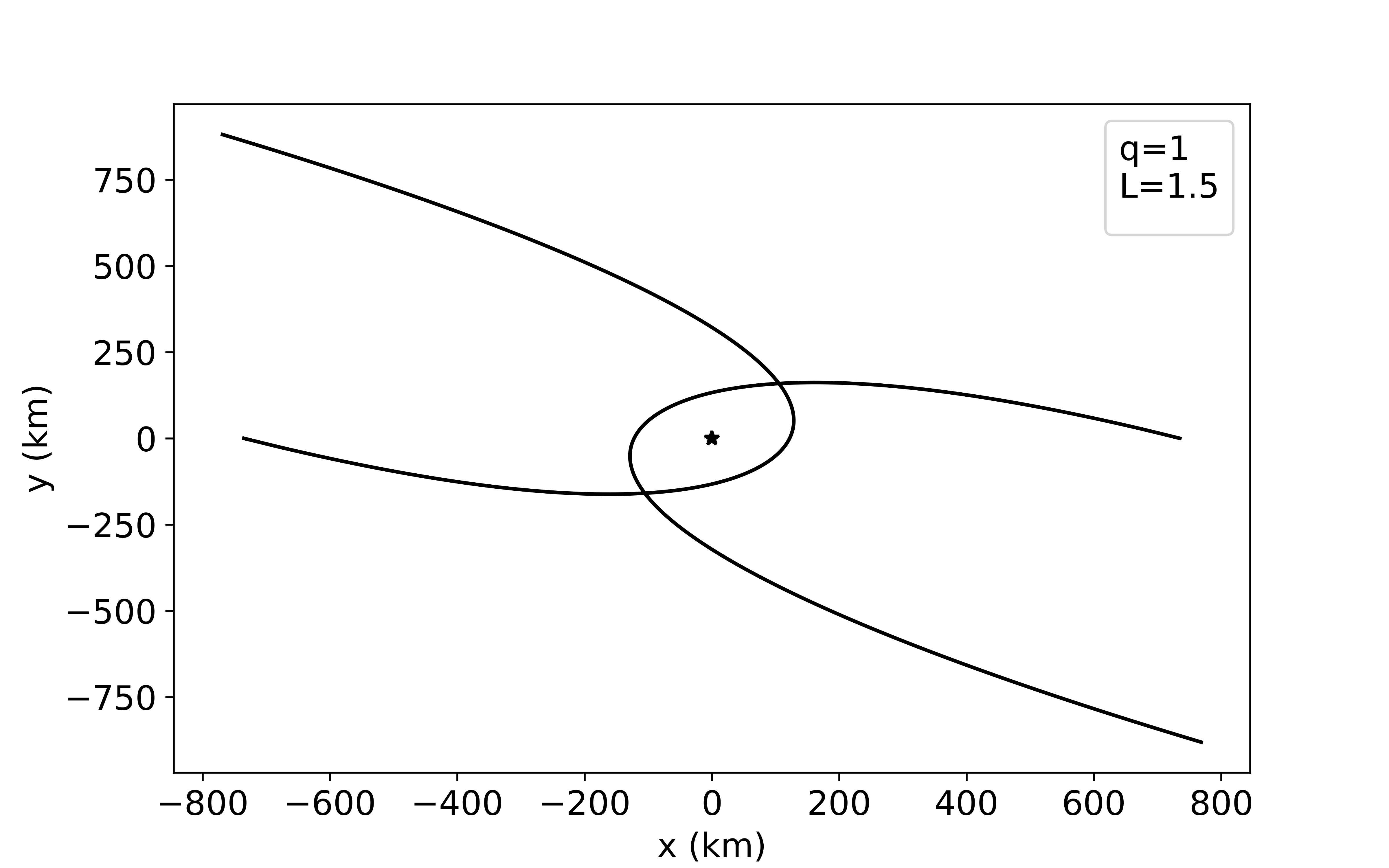} \hfill
\includegraphics[width=.45\textwidth]{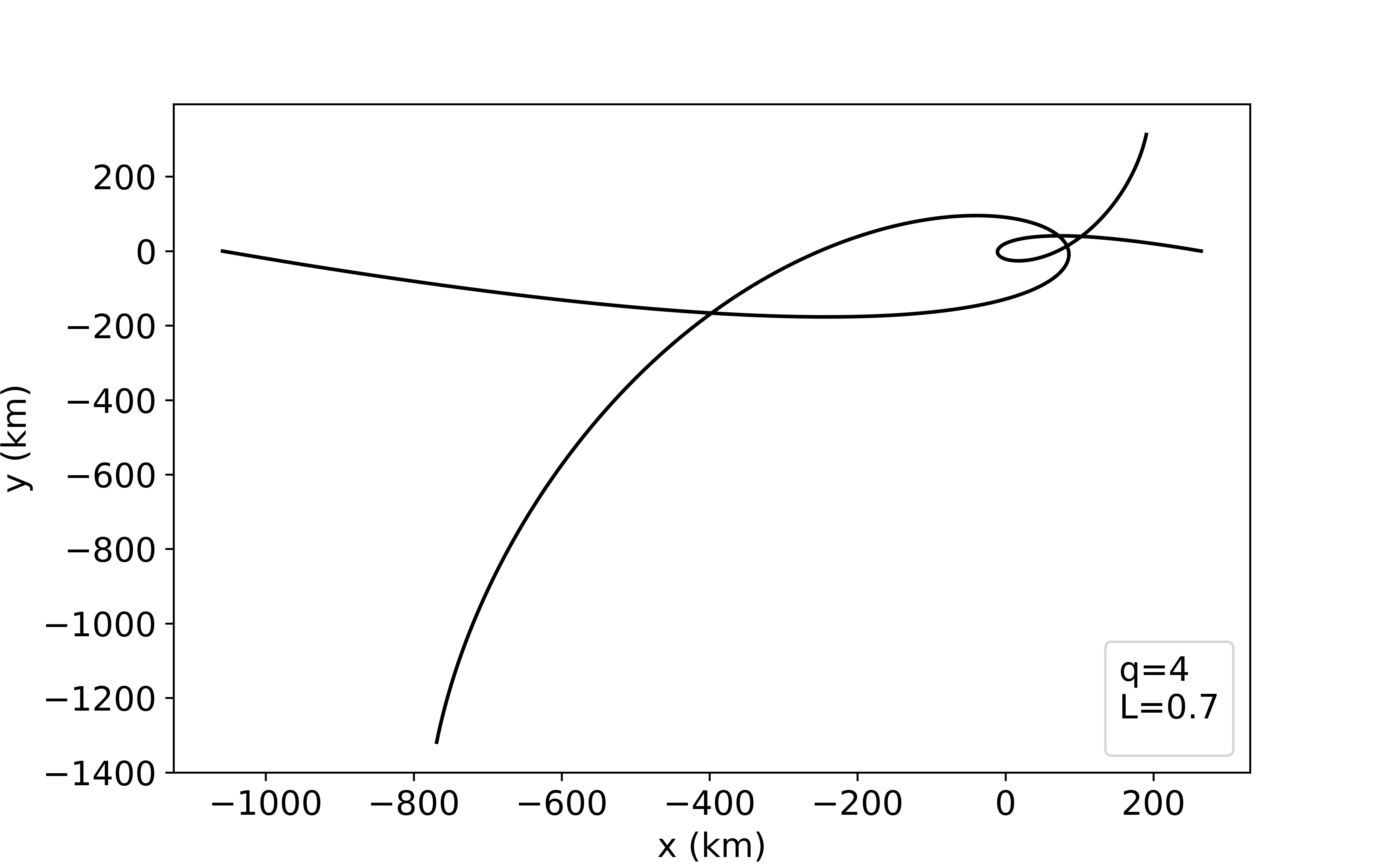}
\caption{Trajectories for three of the waveforms used in this study, with input parameters $(q, L) = (1, 1.2)$ (top), $(q, L) = (1, 1.5) $(center), $(q, L) = (4, 0.7)$ (bottom)}
\label{fig:trajecs}
\end{figure}

We utilized numerical simulations from a study of the critical cross section for dynamical capture of two black holes~\cite{Bae:2017crk}.
The original study mapped energy and angular momentum of hyperbolic orbits over a wide parameter space and simulated parabolic orbits, assuming nearly identical gravitational wave emission due to similar orbital shapes at periapsis.
While parabolic orbits represent the threshold between bound and unbound motion, hyperbolic and parabolic orbits are indistinguishable for burst \gw{} detection purposes.
For a fixed energy, varying the angular momentum is equivalent to considering different impact parameters. 
Based on angular momentum magnitude, orbits can be classified as either direct merging or fly-by. 
Here, we  used the quadrupolar dominant $(l=2,m=2)$ mode in spin-weighted spherical harmonics from the fly-by orbit,  as shown in Fig.~\ref{fig:waveforms}. 
The orbit characteristics vary with the mass ratio of the two black holes and the total angular momentum (see \ref{fig:trajecs}), which in turn affects the wavelength and intensity of the emitted gravitational waves. 
Previous studies~\cite{OLeary:2008myb, Kocsis:2006hq}  demonstrated that hyperbolic encounters of \bbh{} systems are typically nearly parabolic, and signals from strictly hyperbolic orbits do not differ significantly from parabolic orbits when estimating detection rates.  
For simplicity, we considered only binaries with zero inclination (face-on orientation).

%%%%%%%%%%%%%%%%%%%%%%%%%%%%%%%%%%%%%%%%%%%
\subsection{Analysis Setup}
\label{sec:AnalysisSetup}
%%%%%%%%%%%%%%%%%%%%%%%%%%%%%%%%%%%%%%%%%%%

We injected numerical simulations from~\cite{Bae:2017crk} 
representing \gw{} emission from \bbh{} hyperbolic encounters.
Our dataset includes fifteen waveforms spanning mass ratios $q = [1, 2, 4, 8, 16]$.  The initial angular momenta and, consequently, the impact parameters were selected to ensure that the encounters produce sufficient luminosity to be detectable while avoiding dynamical capture leading to coalescence.
These waveforms were injected into simulated Gaussian noise colored to match the power spectral density $S_n (f)$ of an idealized two-detector network. We then used \bw{} to recover and analyze these signals. 
The initial parameters for this simulation study are listed in Table~\ref{tab:det_rates}. 

For each injection, we calculated the signal-to-noise ratio (SNR) to determine the maximum detectable distance and the overlap distribution across iterations, which provides a quantitative measure of the reconstruction's faithfulness to the original signal.

We adopted uniform priors on the frame function parameters, within these  bounds: 
\begin{itemize}
\item time:  $t_{min} \leq t_0 \leq t_{max}$, including a 1 s interval centered on the peak amplitude of the injection;
\item frequency: $ f_{min} \leq f_0 \leq f_{max}$ where $f_{min}$ is set at 20 Hz and $f_{max}$ is the Nyquist frequency (4096 Hz in our case, determined by the sampling rate);
 \item quality factor: $ 0.1 \leq Q \leq 40$;
%\item $Q$ uniform over $[0.1,40]$
\item phase: $ 0 \leq \phi \leq 2 \pi$.
\end{itemize}

\noindent
We use the prior on the number of frame functions as described in Eq.~\ref{eq:wavletPrior}. We conducted our analysis using \bw{} with 4-second segments and 4096 Hz sampling rate, reconstructing each waveform with the three frame functions (Morlet-Gabor, chirplet, and shapelet). This process was repeated for LIGO, A+, CE, and ET power spectral densities, totaling 180 injections.

For each frame function, we calculate the overlap ${\cal{O}}$ between the reconstructed and injected signals, where ${\cal{O}} = 1$ indicates complete agreement. The noise curves used were extracted from the publicly available LIGO sensitivity design curves~\cite{Moore:2014lga} and are shown in Fig.~\ref{fig:PSDs}.

%%%%%%%%%%%%%%%%%%%%%%%%%%%%%%%%%%%%%%%%%%%
\section{Results}
%%%%%%%%%%%%%%%%%%%%%%%%%%%%%%%%%%%%%%%%%%%

We present our main results on the basis of the analytical framework described above. 
First, we evaluate \bw{} reconstruction faithfulness by examining overlap distributions for each frame function and analyzing which waveform characteristics favor shapelets over other methods.
We then calculate detection rates for LIGO, A +, CE, and ET, based on the event rate estimates of Kocsis et al.~\cite{Kocsis:2006hq}, identifying regions of the parameter space with the highest detection probability.

\subsection{Comparison of frame functions}

\begin{figure*} [p] %[hbt]
        \includegraphics[width=\textwidth]{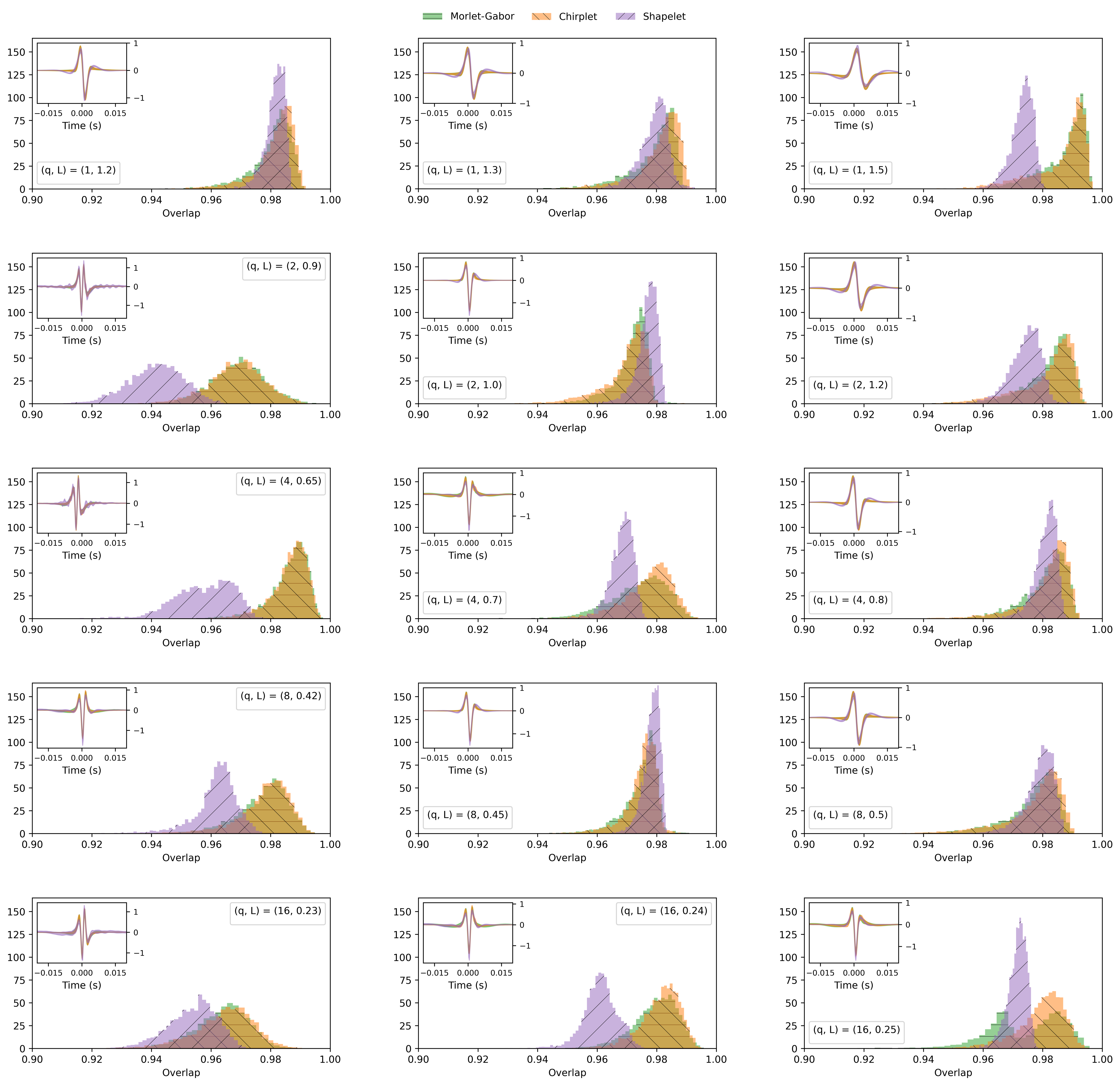}
        \caption{Comparison of waveform reconstructions with three frame functions. In each panel, the inset plot is the 90 percent credible interval whitened waveform reconstruction using the Morlet-Gabor (green), chirplet (orange), and shapelet (purple), with time on the abscissa and dimensionless strain on the ordinate. The main plot in each panel is the posterior distribution of the overlaps between the injected waveform and the Morlet-Gabor, chirplet, and shapelet reconstructions (same colors as the inset plots). One can see that in nearly all cases, the overlaps distribution of chirplet and Morlet-Gabor is nearly identical, owing to their similar parameterizations.}
        \label{fig:hyp-compare}
    \end{figure*}

\begin{figure}[tbh]
\hspace*{-1.5cm}
        \includegraphics[width=0.6\textwidth]{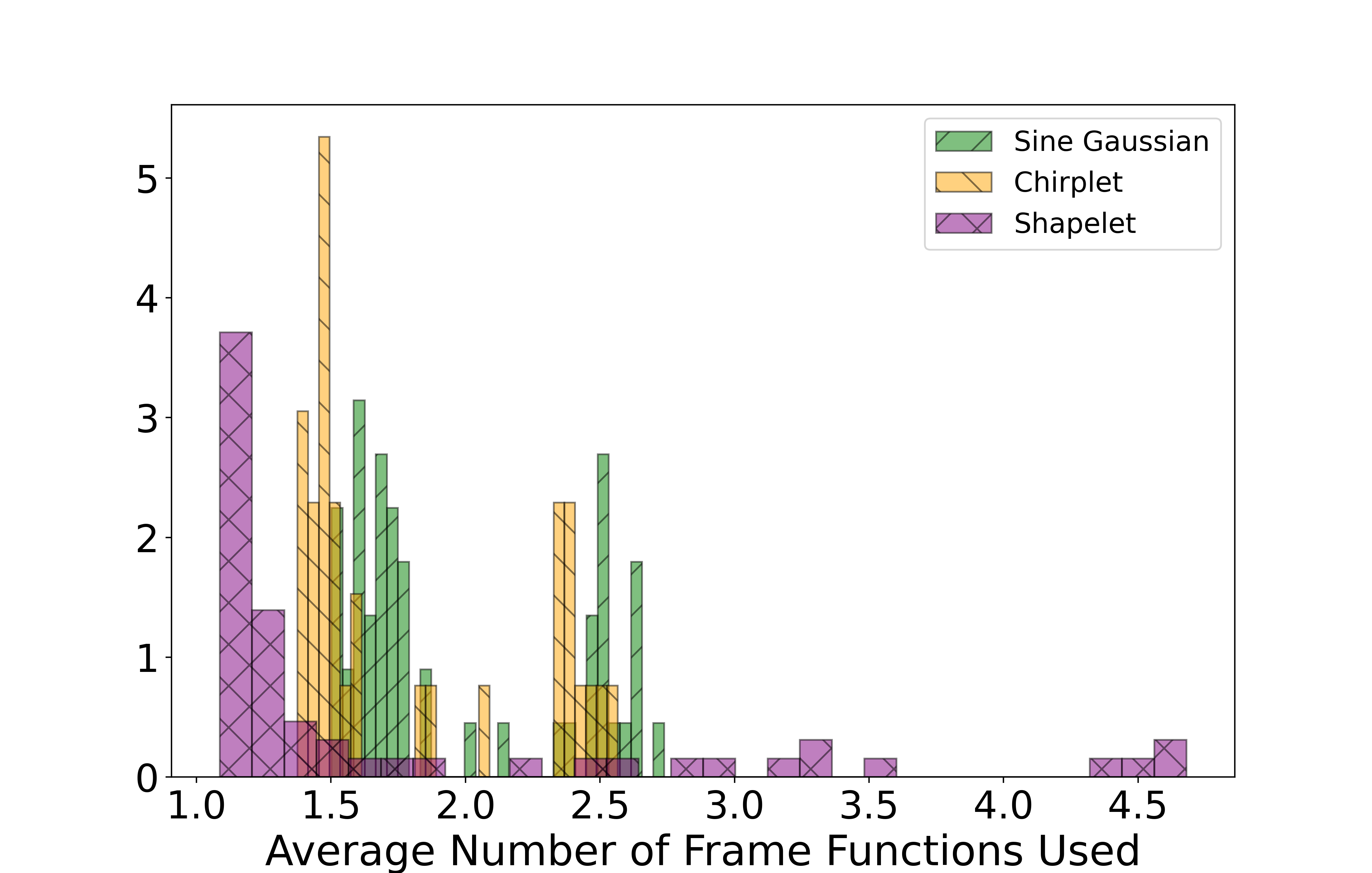}
        \caption{
        Average number of frame functions used in three frame function types—Morlet-Gabor, Chirplet, and Shapelet—for the reconstructions of hyperbolic encounter waveforms using Morlet-Gabor wavelets, chirplets and shapelets. Note that the distributions for Morlet-Gabor and chirplet exhibit double peaks, indicating two parts of the reconstructed parameter space.  %\LC{which hyperbolic waveforms? is this one each of the waveforms in fig 5? or more than one instance? }
        %%%%% Histograms showing the average number of frame functions used in three frame function types—Morlet-Gabor, Chirplet, and Shapelet—for the reconstructions of hyperbolic encounter waveforms, as previously presented in the paper. Note that the distributions for Morlet-Gabor and Chirplet exhibit double peaks, indicating two parts of the reconstructed parameter space.             
        }
        \label{fig:avg_no_basis}
    \end{figure}

{Figure~\ref{fig:hyp-compare} compares  \bw{} reconstruction performance with the three frame functions across hyperbolic encounter parameters. 
Panels are arranged by mass ratio (top to bottom) and initial angular momentum (left to right), showing overlap posteriors and 90\% credible waveform reconstructions. 
%
% MM-- rewriting this section below because "narrower" and "less skewed" posteriors don't necessarily lead to improvement
%Across most of the parameter space, shapelets yield narrower, more sharply peaked overlap distributions than chirplets and Morlet-Gabor wavelets, 
%indicating more stable reconstructions, and suggesting that shapelets are  well-suited to model the localized, burst-like features of hyperbolic encounters. 

%Two configurations, $(q, L) = (2, 0.9)$ and $(q, L) = (4, 0.65)$, exhibit notably lower overlaps across all frame functions. 
%Since the results control for SNR, these outliers correspond to more complex morphologies which challenge the reconstruction. 
%In particular, the $(4, 0.65)$ signal required more basis functions to converge\MM{if you want to include this statement, need more info here}, pointing to limitations from  restricting the shapelet index to $n = 2$. While these cases point to regions of the parameter space with degraded performance, they are exceptions to the overall trend where shapelets perform reliably. 

Across the parameter space, all frame functions produce overlaps greater that 0.9, indicating they can all reconstruct the hyperbolic encounters well.  We see nearly identical overlap distributions for chirplets and wavelets. This result is not unexpected, as chirplets reduce to wavelets when $\dot{f}_0=0$ (see Eq.~\ref{eq:chirplets}).  The distribution of overlaps using shapelets for the reconstruction are noticeably distinct from the chirplets and wavelets results; sometimes the shapelet-based reconstruction produces higher overlaps on average (for example $(q,L)=(2,1.0)$), and sometimes lower overlaps on average (for example $(q,L)=(4,0.65)$). This potentially points to limitations from  restricting the shapelet index to $n = 2$, and future work can investigate changing that index.

%The selection of frame functions balances computational cost with reconstruction accuracy. 
%Despite this, shapelets produce narrower, less skewed posterior overlap distributions, which may improve the consistency and reliability of signal reconstruction~\cite{torrence1998practical, mallat1999wavelet, PhysRevResearch.4.033078}. 
%\LC{these references are for shapelet theory, while "posterior overlap distributions" refers to the plots in  figure 7} \PL{Added more relevant citations}

% MM -- I think Sec II B covers this stuff about chirplets suiting chirp like signals so it's just repitition here. Commenting out.
%Chirplets suit chirp-like signals, such as binary black hole mergers, due to their adaptability to time-varying frequencies~\cite{Millhouse:2018dgi, Henshaw:2024lza}.
%\MM{chirplets are still about 1.5x slower than wavelets}\PL{I removed the comment on computational efficiency}. 
%Morlet-Gabor wavelets are simpler, making them particularly useful for constant-frequency signals~\cite{Cornish:2014kda, Henshaw:2024lza}. 

In addition to reconstruction fidelity, we also investigate the typical model dimensionality for the three frame functions. Figure~\ref{fig:avg_no_basis} shows the average (mean) number of frame functions used per BayesWave run for each hyperbolic waveform described in Sec.~\ref{sec:AnalysisSetup}, injected with SNRs uniformly distributed from 10 to 40. 
%\MM{Sorry to belabor this point again, but I do think it's necessary to spell out exactly what Fig 8 is showing.  I thought it showed the average number of frame functions used in an MCMC run on each of the 15 waveforms shown in fig. 7. However the histogram has more than 15 counts, so I still don't know what that figure is showing.  Do you maybe have a pointer to how you compiled these the data to make this histogram?}. 
%Morlet-Gabor and chirplet wavelets typically require fewer frame functions, generally capping around four.  %MM-- editing this because I don't think it's true to say chirplets and wavelets "typically require fewer"
%\PL{The data from the plot originally was meant to represent the average number of frame functions per iteration as a function of SNR, which ranges from 10 to 40. So that's why there are more than 15 bins. The double-peakedness could indeed arise from the fact that some hyperbolic encounter signals are approximately symmetric about $t = t_0$, while others are not, which can lead to variation in the number of frame functions required to represent them. We can add this into the text.}
While the number of shapelets used peaks at a lower value (close to one) compared to wavelets and chirplets, the tail of the shapelets distribution extends to higher values. This indicates that while chirplets and wavelets provide consistently high overlaps, shapelets sometimes achieve higher or lower overlaps depending on the waveform, and the broader tail of their distribution suggests that more components may occasionally be required, implying that shapelets could be chosen when capturing complex waveform features is a priority, while chirplets or wavelets may be preferred for efficiency in simpler cases.
%The chirplet and wavelet distributions also exhibit distinct double peaks, likely reflecting the MCMC algorithm exploring two reconstruction regimes: one favoring simpler models with fewer wavelets, another allowing increased complexity to capture additional signal details\MM{This is conflating different aspects of what the plot is showing.  Unless I still don't understand how the histograms are being produced, each count in a bin is a different MCMC run on a different signal.  So you can't say the MCMC is exploring different regimes since it's just presenting a single number from each MCMC run. My guess is the double-peakedness is related to how some hyperbolic signals are symmetric about t=t0 and some are not}. This reflects the balance between model sparsity and accurate signal representation.

We additionally compare the computational efficiencies of the different frame functions. Table~\ref{tab:disk-usage} shows that shapelets use less disk space but require more memory and CPU time than  Morlet-Gabor and chirplet frames. 
% MM -- commenting this out because it contradicts what's shown earlier (shapelets tend to have lwoer overlaps...)
%Shapelets demonstrate superior reconstruction performance due to their flexible, adaptive shapes better suited for capturing complex or localized features in the signal. However, this flexibility sometimes leads to selecting more frame functions, occasionally five or six, when subtle or irregular structures are present that simpler wavelets cannot efficiently represent. While this results in higher complexity, it also enables a more precise and faithful signal reconstruction in challenging cases.

\subsection{Detection rates}

In this study, we declare that a detection occurs when the recovered signal SNR exceeds 8, consistent with other similar work~\cite{Cahillane:2022pqm}. 
Since the SNR is proportional to the waveform amplitude $h$, (see Eq.~\ref{eq:inner_prod}), it is  inversely proportional to the (horizon) luminosity distance $d_L$, the largest distance at which the signal is detectable:
\begin{equation}
    \text{SNR} \sim \frac{1}{d_L}
\end{equation}

For non-negligible redshift ($z \approx 0.05$) we must consider the rate of cosmological expansion, using the \textit{redshifted sensitivity volume}:
\begin{equation}
    V_z = 4 \pi \int^{z_{max}}_{0}  (1+z)\frac{d V_C}{d z} dz,
\end{equation}
where $z_{max}$ is defined as the maximum redshift at which a source would produce an SNR of exactly 8 in the detector, $\frac{d V_C}{d z}$ is the \textit{differential comoving volume}:
\begin{equation}
    \frac{d V_C}{d z} = 4 \pi \frac{d^2_L}{E (z) (1 + z)^2}
\end{equation}
and $E (z) = \sqrt{\Omega_{\Lambda} + \Omega_{m}(1 + z)^3}$. $\Omega_{\Lambda}$ and $\Omega_{m}$ are the fractions of mass-energy contained in vacuum energy, and matter, respectively.

The detection rate $\nu$ for a given waveform is related to the sensitivity volume by:
\begin{equation}
    \nu = {\cal{R}} V,
\end{equation}
with ${\cal{R}}$ the expected rate for hyperbolic encounters in globular clusters. For all waveforms, we use the expected rate derived by Kocsis et al.~\cite{Kocsis:2006hq}.

\begin{table*}[tb]
\caption{Tabulated results from numerical injections.  The first and second columns are the mass ratio and the initial angular momentum in geometric units, respectively, i.e. the initial parameters of the \bbh{} system corresponding to each injected waveform. The third through tenth columns are the sensitive luminosity distance and detection rate for LIGO, A+, CE, and ET, respectively.}
\label{tab:det_rates}

\begin{ruledtabular}
\begin{tabular}{c|c|cc|cc|cc|cc}
%
% &\multicolumn{2}{c}{$D_{4h}^1$}&\multicolumn{2}{c}{$D_{4h}^5$}\\
% Ion&1st alternative&2nd alternative&lst alternative
%&2nd alternative\\ \hline
% K&$(2e)+(2f)$&$(4i)$ &$(2c)+(2d)$&$(4f)$ \\
% Mn&$(2g)$\footnote{The $z$ parameter of these positions is $z\sim\frac{1}{4}$.}
% &$(a)+(b)+(c)+(d)$&$(4e)$&$(2a)+(2b)$\\
% Cl&$(a)+(b)+(c)+(d)$&$(2g)$\footnotemark[1]
% &$(4e)^{\text{a}}$\\
% He&$(8r)^{\text{a}}$&$(4j)^{\text{a}}$&$(4g)^{\text{a}}$\\
% Ag& &$(4k)^{\text{a}}$& &$(4h)^{\text{a}}$\\
%
\multirow{2}{*}{q}    & \multirow{2}{*}{L}    & \multicolumn{2}{c|}{LIGO} & \multicolumn{2}{c|}{ LIGO A+}  & \multicolumn{2}{c|}{ Cosmic Explorer} & \multicolumn{2}{c}{ Einstein Telescope}\\
    &     &    D [Mpc] & Rate [$10^{-3}$/y] & [Mpc] & Rate [$10^{-3}$/y] & [Mpc] &  Rate [y$^{-1}$] & [Mpc] & Rate [y$^{-1}$]\\
\hline
\multirow{3}{*}{1}    & 1.2 & 165   & 1.24 & 305  & 7.47 & 2676    &  2.66 & 1715  &   0.86    \\
    & 1.3 & 141 & 0.800 & 256  &  4.47 & 2623    & 2.52 & 1470   &   0.58  \\
    & 1.5 & 99 & 0.277 & 181 & 1.64 & 2116     & 1.48 & 1050  &   0.24   \\
\hline
\multirow{3}{*}{2}     & 0.9 & 197 & 2.09 & 372 & 13.3 & 3081    & 3.75 & 1969   &  1.23   \\
    & 1.0  & 164 & 1.22 & 304 & 7.33 & 2634    & 2.55 & 1720    &  0.87  \\
    & 1.2  & 115 & 0.441 & 214  & 2.64 & 2272    & 1.77 & 1177    &  0.32  \\
\hline
\multirow{3}{*}{4}     & 0.65 & 144 & 0.851 & 291 & 6.64 & 2237    & 1.70 & 1458    &   0.57   \\
    & 0.7  & 124    & 0.537 & 231  & 3.30 & 2040    & 1.35 & 1257   &   0.38  \\
 & 0.8  & 97 & 0.260 & 184 & 1.70 & 1723    & 0.87 & 989  &     0.20 \\
 \hline
\multirow{3}{*}{8}     & 0.42 & 80 & 0.147 & 153 & 0.991 & 1260   &  0.39 & 828  &    0.12   \\
    & 0.45 & 71 & 0.103 & 132  & 0.646 & 1169    & 0.32 & 746  &   0.09   \\
    & 0.5  & 58 & 0.056 & 110 & 0.380 & 1024      & 0.22 & 597   &   0.05  \\
\hline
\multirow{3}{*}{16}    & 0.23 & 49    & 0.032 & 94.1  & 0.237 & 762   & 0.098 & 497   &  0.03   \\
   & 0.24 & 45 & 0.027 & 87 & 0.188 & 728    & 0.087 & 442   &    0.02 \\
   & 0.25 & 41    & 0.021 & 78 & 0.137 & 687    & 0.074 & 429    &    0.02
\end{tabular}
\end{ruledtabular}
\end{table*}
%\LC{need to elaborate here on how these constraints were obtained. }
\begin{figure*}[hbt]
\centering
        \includegraphics[width=\textwidth]{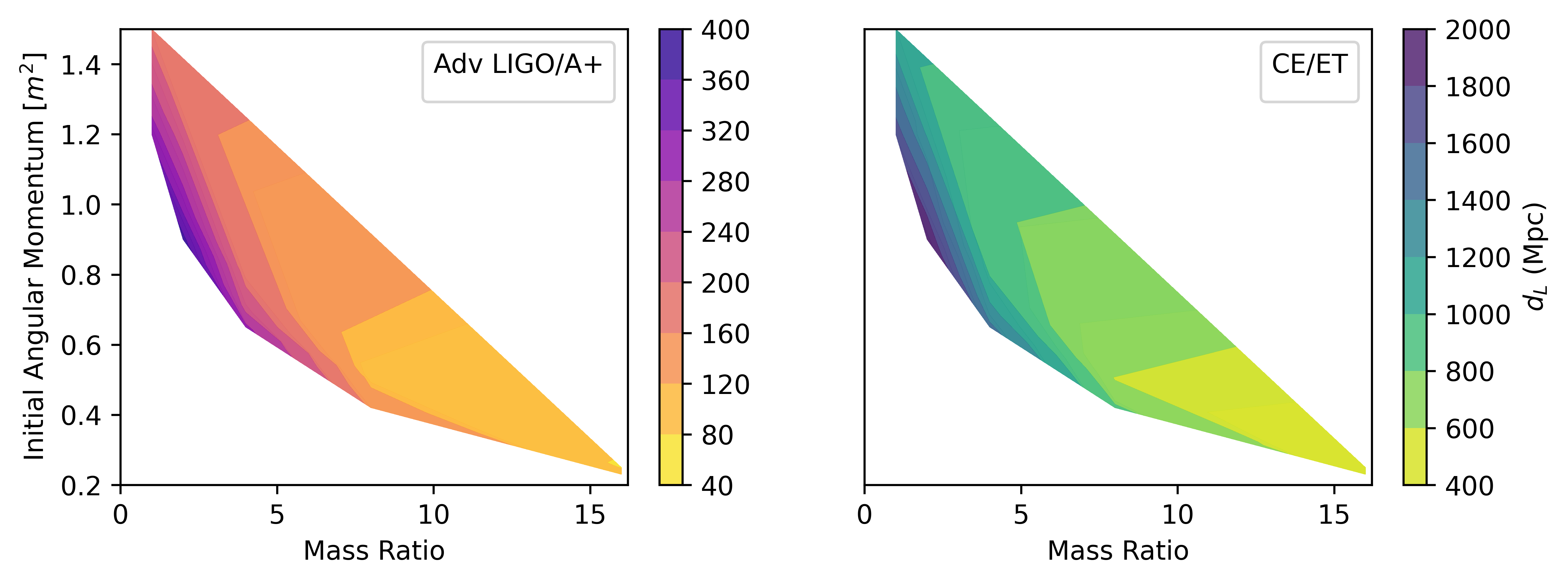}
        \caption{Sensitivity distance contours from our waveform injections. For both panels, the ordinate is the initial angular momentum in geometric units of $m^2$, and the abscissa is the mass ratio. The color bar shown to the right of each panel shows the luminosity distances at which the injection is detectable for SNR threshold 8, given an idealized noise spectrum. The left panel is for LIGO and A+ and the right panel is for CE and ET. Although our parameter space is limited, it is clear that the equal-mass ratio provides detectability at greater luminosity distance compared to the more extreme mass ratio counterparts, which is sensible because the gravitational radiation from bremsstrahlung encounters is proportional to the mass. The sensitivity range of $40 - 400$ Mpc is consistent with estimates by~\cite{Kocsis:2006hq}, who examined an equal mass $m_1 = m_2 = 40\,\msol$ case and found an estimated range of $200$ Mpc for a threshold of SNR 5 studied the LIGO and A+ detector cases. Similarly, ~\cite{Bini:2023gaj} found luminosity distances $40-50$ Mpc range. The estimates presented here are comparatively optimistic, owing to the choice of idealized noise (i.e. no glitches).}
        %\LC{missing caption}
        \label{fig:contour_dist}
    \end{figure*}

\begin{table*}[tb]
\caption{Approximate average disk memory usage and CPU time in reconstruction analysis. }
\label{tab:disk-usage}
\begin{tabular}{|l|l|l|l|}
\hline
Frame Function & Disk Usage (MB) & Memory Usage (MB) & CPU Time (s) \\ \hline
Morlet-Gabor & 11.179          & 128               & 6172             \\ \hline
Chirplet     & 14.474          & 129               & 6175             \\ \hline
Shapelet     & 9.16            & 150               & 10991            \\ \hline

\end{tabular}
\end{table*}

Table~\ref{tab:det_rates} reports the calculated luminosity distance constraints and  detection rates for \bbh{} hyperbolic encounters, using the injection set from Section III, assuming idealized noise and optimal sky orientation for a LIGO-type detector; our results are therefore optimistic. Actual noise (thermal, electronic, seismic) can distort the signals and reduce their detectability, while non-optimal detector orientation and systematic errors introduce additional limitations and biases that are not captured under ideal conditions.
We repeated this analysis for A+, Cosmic Explorer, and Einstein Telescope noise realizations. 

Our order of magnitude luminosity sensitivity distances are comparable to previous calculations~\cite{10.1093/mnras/stab2721, Kocsis:2006hq, Bini:2023gaj}, with \ligo{} sensitivity distance ranging from 60 to 250 Mpc, consistent with previous studies~\cite{Kocsis:2006hq, 10.1093/mnras/stab2721, Bini:2023gaj}; 
Fig.~\ref{fig:contour_dist} shows sensitivity distance contours as a function of the initial angular momentum (i.e., impact parameter) and the mass ratio. 

The calculated rates suggest that these events are most likely detactable by the next generation of gravitational wave detectors, Cosmic Explorer and Einstein Telescope, with 2.66 events per year for equal mass cases with Cosmic Explorer. Next-generation detector rates are overall consistent with other studies~\cite{10.1093/mnras/stab2721}.

%%%%%%%%%%%%%%%%%%%%%%%%%%%%%%%%%%%%%%%%%%%
\section{Conclusion}  
%%%%%%%%%%%%%%%%%%%%%%%%%%%%%%%%%%%%%%%%%%%

Highly eccentric black hole binary systems are predicted to be common in globular clusters and dense astrophysical environments~\cite{10.1093/mnras/stab2721, Kocsis:2006hq, Capozziello:2008mn, Morras:2021atg, Li2022, Caldarola:2023ipo}. In particular,
\bbh{} hyperbolic encounters of stellar-mass black holes should emit \gw{s} manifesting as single-cycle transients in current ground-based detector frequency ranges~\cite{Kocsis:2006hq, Capozziello:2008mn}.

This paper presents a Bayesian analysis of numerical waveforms  from \bbh{} hyperbolic encounters and estimates detection rates for current and future ground-based interferometers. 
Our rates are consistent with previous work~\cite{Kocsis:2006hq, 10.1093/mnras/stab2721, Bini:2023gaj}, $\sim 10^{-3}$ per year for LIGO and A+ and 
$\sim$ few per year for Cosmic Explorer and Einstein Telescope.

%Of the more than ninety \gw{} signals confirmed by ground-based detections cited in the third gravitational-wave catalog~\cite{KAGRA:2021vkt}  to date, all are associated with binary compact objects in circular and quasi-circular orbits (i.e. $e < 0.03$). Therefore, the calculated rates are subject to uncertainties associated with the population of binaries in unbound orbits observable by ground-based detectors.  As current detectors become more sensitive, these rates will become more constrained, which will allow a higher chance of observation. 

The analysis was performed with \bw{}. We tested shapelet frame functions against traditional  Morlet-Gabor and chirplet functions, using the overlap between injected $h_{inj}$ and recovered $h_{rec}$ waveforms
We found that \bw{} with shapelets provides comparable reconstruction quality while typically requiring fewer basis functions, possibly due to morphological similarity between shapelets and \bbh{} hyperbolic encounters in the time-frequency domain. The computational requirements are tabulated in Table~\ref{tab:disk-usage}.

Applying this analysis to real data will require addressing instrumental glitches, particularly blip glitches that morphologically resemble hyperbolic encounter waveforms and could cause false detections. \bw{}'s Bayesian framework for distinguishing astrophysical signals from glitches may directly apply to hyperbolic encounters.

%It will be useful to apply the same analysis-full scale to noisy data, as some instrumental glitches in interferometric data bear a similarity to the waveform in the time domain. Replacing the Gaussianity of the noise spectrum assumed in this analysis with real data containing glitches will negatively impact sensitivity, particularly in the case of blip glitches, which have a morphological similarity to \gw{s} from hyperbolic encounters in the time-frequency domain, possibly leading to false detections. Therefore, rejecting glitches as events would be part of the goal in establishing false alarm rates. Since \bw{} incorporates Bayesian statistics to establish the preference of one model over another (i.e. whether the data contain an astrophysical signal or glitch), this may be directly applied to the case of hyperbolic encounter waveforms for blip glitches.

Given globular cluster models~\cite{Kocsis:2006hq} and the expected increasing detector sensitivity, \gw{s} from hyperbolic encounters are likely to be detectable in the future. Since dynamical evolution scenarios predict eccentric binary signatures~\cite{Maggiore:2019uih}, confirmed detections of hyperbolic encounters may inform black hole formation scenarios.

%Given the globular cluster models used in this study~\cite{Kocsis:2006hq} and the increasing sensitivity of current and future detectors, it means that \gw{s} from hyperbolic encounters are likely to be detected in the future. As current and future gravitational wave detectors become increasingly sensitive, they will be better able to detect signals at lower frequencies~\cite{Reitze:2019iox}.

%Since dynamical evolution scenarios would have signatures of eccentric binaries~\cite{Maggiore:2019uih}, confirmed detections of \gw{s} from hyperbolic encounters and other highly eccentric phenomena may inform the veracity of the dynamical evolution scenario for black hole formation. 

%%%%%%%%%%%%%%%%%%%%%%%%%%%%%%%%%%%%%%%%%%%
\section{Acknowledgment}  
%%%%%%%%%%%%%%%%%%%%%%%%%%%%%%%%%%%%%%%%%%%

This material is based upon work supported by NSF's LIGO Laboratory which is a major facility fully funded by the National Science Foundation. This research utilized data, software, and tools provided by the Gravitational Wave Open Science Center (https://www.gw-openscience.org), a resource maintained by LIGO Laboratory, the LIGO Scientific Collaboration, and the Virgo Collaboration. LIGO's operations are funded by the U.S. National Science Foundation, while Virgo receives support from the French Centre National de Recherche Scientifique (CNRS), the Italian Istituto Nazionale di Fisica Nucleare (INFN), and the Dutch Nikhef, with additional contributions from Polish and Hungarian institutions. The authors also acknowledge the computational resources provided by the LIGO Laboratory, supported by NSF Grants PHY-0757058 and PHY-0823459.
This work was carried out using services offered by the OSG Consortium [59–62], funded by NSF awards 2030508 and 1836650. The GT authors express their gratitude to the NSF for funding under Grants PHY-1809572 and PHY-2110481.
Y.-B.B. was supported in part by the National Research Foundation of Korea (NRF) grants funded by the Korea government (MSIT) (RS-2025-00556091 and RS-2025-00564350).
The authors acknowledge John Wise, Daniel Williams, and Zoltan Haiman for illuminating discussions.
\bibliography{reference}
\end{document}

%% file: def.tex
\def\gt#1{Georgia Institute of Technology#1 (GeorgiaTech#1)\gdef\gt{GeorgiaTech}}
\def\si#1{Senior Investigator#1 (SI#1)\gdef\si{SI}}
\def\emri#1{Extreme Mass-Ratio Inspiral#1 (EMRI#1)\gdef\emri{EMRI}}
\def\imbh#1{Intermediate Mass Black Hole#1 (IMBH#1)\gdef\imbh{IMBH}}
\def\smbh#1{supermassive black hole#1(SMBH#1)\gdef\smbh{SMBH}}
\def\bbh#1{binary black hole#1 (BBH#1)\gdef\bbh{BBH}}
\def\pbh#1{primordial black hole#1 (PBH#1)\gdef\pbh{PBH}}
\def\imbhb#1{intermediate mass black hole binary#1 (IMBHB#1)\gdef\imbhb{IMBHB}}
\def\hmns#1{hypermassive neutron star#1 (HMNS#1)\gdef\hmns{HMNS}}
\def\bh#1{black hole#1 (BH#1)\gdef\bh{BH}}
\def\ns#1{neutron star#1 (NS#1)\gdef\ns{NS}}
\def\hmns#1{hyper-massive neutron star#1 (HMNS#1)\gdef\hmns{HMNS}}
\def\nsbh#1{neutron star-black hole#1 (NSBH#1)\gdef\nsbh{NSBH}}
\def\bns#1{binary neutron star#1 (BNS#1)\gdef\bns{BNS}}
\def\gw#1{gravitational wave#1 (GW#1)\gdef\gw{GW}}
\def\eos#1{equation of state#1 (EOS#1)\gdef\eos{EOS}}
\def\gpu#1{graphics processing unit#1 (GPU#1)\gdef\gpu{GPU}}
\def\gr#1{General Relativity#1 (GR#1)\gdef\gr{GR}}
\def\cbc#1{Compact Binary Coalescence#1 (CBC#1)\gdef\cbc{CBC}}
\def\eob#1{Effective-One-Body#1 (EOB#1)\gdef\eob{EOB}}
\def\pnw#1{post-Newtonian#1 (PN#1)\gdef\pnw{PN}}
\def\pmw#1{post-Minkowskian#1 (PM#1)\gdef\pmw{PM}}
\def\hom#1{Higher Order Mode#1 (HOM#1)\gdef\hom{HOM}}
\def\ligo#1{Laser Interferometer Gravitational-Wave Observatory#1 (LIGO#1)\gdef\ligo{LIGO}}
\def\lisa#1{Laser Interferometer Space Antennae#1 (LISA#1)\gdef\lisa{LISA}}
\newcommand{\bw}{{BayesWave}}

\newcommand{\msol}{{\mathrm{M}_{\odot}}}

\usepackage{xcolor}
\usepackage{comment}
\usepackage[normalem]{ulem}
\usepackage{soul}   % provides \hl{} for highlighting

\def\gt#1{Georgia Institute of Technology#1 (GeorgiaTech#1)\gdef\gt{GeorgiaTech}}
\def\si#1{Senior Investigator#1 (SI#1)\gdef\si{SI}}
\def\emri#1{Extreme Mass-Ratio Inspiral#1 (EMRI#1)\gdef\emri{EMRI}}
\def\imbh#1{Intermediate Mass Black Hole#1 (IMBH#1)\gdef\imbh{IMBH}}
\def\smbh#1{supermassive black hole#1(SMBH#1)\gdef\smbh{SMBH}}
\def\bbh#1{Binary Black Hole#1 (BBH#1)\gdef\bbh{BBH}}
\def\imbhb#1{intermediate mass black hole binary#1 (IMBHB#1)\gdef\imbhb{IMBHB}}
\def\hmns#1{hypermassive neutron star#1 (HMNS#1)\gdef\hmns{HMNS}}
\def\bh#1{black hole#1 (BH#1)\gdef\bh{BH}}
\def\ns#1{neutron star#1 (NS#1)\gdef\ns{NS}}
\def\hmns#1{hyper-massive neutron star#1 (HMNS#1)\gdef\hmns{HMNS}}
\def\nsbh#1{neutron star-black hole#1 (NSBH#1)\gdef\nsbh{NSBH}}
\def\bns#1{binary neutron star#1 (BNS#1)\gdef\bns{BNS}}
\def\gw#1{gravitational wave#1 (GW#1)\gdef\gw{GW}}
\def\eos#1{equation of state#1 (EOS#1)\gdef\eos{EOS}}
\def\gpu#1{graphics processing unit#1 (GPU#1)\gdef\gpu{GPU}}
\def\gr#1{General Relativity#1 (GR#1)\gdef\gr{GR}}
\def\cbc#1{Compact Binary Coalescence#1 (CBC#1)\gdef\cbc{CBC}}
\def\eob#1{Effective-One-Body#1 (EOB#1)\gdef\eob{EOB}}
\def\pnw#1{post-Newtonian#1 (PN#1)\gdef\pnw{PN}}
\def\pmw#1{post-Minkowskian#1 (PM#1)\gdef\pmw{PM}}
\def\hom#1{Higher Order Mode#1 (HOM#1)\gdef\hom{HOM}}
\def\ligo#1{Laser Interferometer Gravitational-Wave Observatory#1 (LIGO#1)\gdef\ligo{LIGO}}
\def\lisa#1{Laser Interferometer Space Antennae#1 (LISA#1)\gdef\lisa{LISA}}

%% file: main-1124.bbl
\begin{thebibliography}{77}
\providecommand{\natexlab}[1]{#1}
\providecommand{\url}[1]{\texttt{#1}}
\expandafter\ifx\csname urlstyle\endcsname\relax
  \providecommand{\doi}[1]{doi: #1}\else
  \providecommand{\doi}{doi: \begingroup \urlstyle{rm}\Url}\fi

\bibitem[Abbott et~al.(2021{\natexlab{a}})]{LIGOScientific:2021djp}
R.~Abbott et~al.
\newblock {GWTC-3: Compact Binary Coalescences Observed by LIGO and Virgo
  During the Second Part of the Third Observing Run}, 11 2021{\natexlab{a}}.
\newblock {Preprint arXiv:2111.03606}.

\bibitem[Abac et~al.(2025)]{LIGOScientific:2025slb}
A.~G. Abac et~al.
\newblock {GWTC-4.0: Updating the Gravitational-Wave Transient Catalog with
  Observations from the First Part of the Fourth LIGO-Virgo-KAGRA Observing
  Run}.
\newblock 8 2025.

\bibitem[Abbott et~al.(2021{\natexlab{b}})]{LIGOScientific:2021qlt}
R.~Abbott et~al.
\newblock {Observation of Gravitational Waves from Two Neutron
  Star\textendash{}Black Hole Coalescences}.
\newblock \emph{Astrophys. J. Lett.}, 915\penalty0 (1):\penalty0 L5,
  2021{\natexlab{b}}.
\newblock \doi{10.3847/2041-8213/ac082e}.

\bibitem[Abbott et~al.(2017)]{LIGOScientific:2017vwq}
B.~P. Abbott et~al.
\newblock {GW170817: Observation of Gravitational Waves from a Binary Neutron
  Star Inspiral}.
\newblock \emph{Phys. Rev. Lett.}, 119\penalty0 (16):\penalty0 161101, 2017.
\newblock \doi{10.1103/PhysRevLett.119.161101}.

\bibitem[Abbott et~al.(2020{\natexlab{a}})]{LIGOScientific:2020aai}
B.~P. Abbott et~al.
\newblock {GW190425: Observation of a Compact Binary Coalescence with Total
  Mass $\sim 3.4 M_{\odot}$}.
\newblock \emph{Astrophys. J. Lett.}, 892\penalty0 (1):\penalty0 L3,
  2020{\natexlab{a}}.
\newblock \doi{10.3847/2041-8213/ab75f5}.

\bibitem[Abbott et~al.(2020{\natexlab{b}})]{LIGOScientific:2020zkf}
R.~Abbott et~al.
\newblock {GW190814: Gravitational Waves from the Coalescence of a 23 Solar
  Mass Black Hole with a 2.6 Solar Mass Compact Object}.
\newblock \emph{Astrophys. J. Lett.}, 896\penalty0 (2):\penalty0 L44,
  2020{\natexlab{b}}.
\newblock \doi{10.3847/2041-8213/ab960f}.

\bibitem[Reitze et~al.(2019)]{Reitze:2019iox}
D.~Reitze et~al.
\newblock {Cosmic Explorer: The U.S. Contribution to Gravitational-Wave
  Astronomy beyond LIGO}.
\newblock \emph{Bull. Am. Astron. Soc.}, 51\penalty0 (7):\penalty0 035, 2019.

\bibitem[Maggiore et~al.(2020)]{Maggiore:2019uih}
M.~Maggiore et~al.
\newblock {Science Case for the Einstein Telescope}.
\newblock \emph{JCAP}, 03:\penalty0 050, 2020.
\newblock \doi{10.1088/1475-7516/2020/03/050}.

\bibitem[Mukherjee et~al.(2021)Mukherjee, Mitra, and
  Chatterjee]{10.1093/mnras/stab2721}
Sajal Mukherjee, Sanjit Mitra, and Sourav Chatterjee.
\newblock {Gravitational wave observatories may be able to detect hyperbolic
  encounters of black holes}.
\newblock \emph{Monthly Notices of the Royal Astronomical Society},
  508\penalty0 (4):\penalty0 5064--5073, 09 2021.
\newblock ISSN 0035-8711.
\newblock \doi{10.1093/mnras/stab2721}.
\newblock URL \url{https://doi.org/10.1093/mnras/stab2721}.

\bibitem[Caldarola et~al.(2024)Caldarola, Kuroyanagi, Nesseris, and
  Garcia-Bellido]{Caldarola:2023ipo}
Marienza Caldarola, Sachiko Kuroyanagi, Savvas Nesseris, and Juan
  Garcia-Bellido.
\newblock {Effects of orbital precession on hyperbolic encounters}.
\newblock \emph{Phys. Rev. D}, 109\penalty0 (6):\penalty0 064001, 2024.
\newblock \doi{10.1103/PhysRevD.109.064001}.

\bibitem[Morr\'as et~al.(2022)Morr\'as, Garc\'\i{}a-Bellido, and
  Nesseris]{Morras:2021atg}
Gonzalo Morr\'as, Juan Garc\'\i{}a-Bellido, and Savvas Nesseris.
\newblock {Search for black hole hyperbolic encounters with gravitational wave
  detectors}.
\newblock \emph{Phys. Dark Univ.}, 35:\penalty0 100932, 2022.
\newblock \doi{10.1016/j.dark.2021.100932}.

\bibitem[Bini et~al.(2023)Bini, Tiwari, Xu, Smith, Ebersold, Principe, Haney,
  Jetzer, and Prodi]{Bini:2023gaj}
Sophie Bini, Shubhanshu Tiwari, Yumeng Xu, Leigh Smith, Michael Ebersold,
  Giacomo Principe, Maria Haney, Philippe Jetzer, and Giovanni~A. Prodi.
\newblock {Search for hyperbolic encounters of compact objects in the third
  LIGO-Virgo-KAGRA observing run}, 2023.
\newblock Preprint arXiv:2311.06630.

\bibitem[Capozziello and De~Laurentis(2008)]{Capozziello:2008mn}
Salvatore Capozziello and Mariafelicia De~Laurentis.
\newblock {Gravitational waves from stellar encounters}.
\newblock \emph{Astropart. Phys.}, 30:\penalty0 105--112, 2008.
\newblock \doi{10.1016/j.astropartphys.2008.07.005}.

\bibitem[Ebersold et~al.(2022)Ebersold, Tiwari, Smith, Bae, Kang, Williams,
  Gopakumar, Heng, and Haney]{Ebersold:2022zvz}
Michael Ebersold, Shubhanshu Tiwari, Leigh Smith, Yeong-Bok Bae, Gungwong Kang,
  Daniel Williams, Achamveedu Gopakumar, Ik~Siong Heng, and Maria Haney.
\newblock {Observational limits on the rate of radiation-driven binary black
  hole capture events}.
\newblock \emph{Phys. Rev. D}, 106\penalty0 (10):\penalty0 104014, 2022.
\newblock \doi{10.1103/PhysRevD.106.104014}.

\bibitem[Garc\'\i{}a-Bellido et~al.(2022)Garc\'\i{}a-Bellido, Jaraba, and
  Kuroyanagi]{Garcia-Bellido:2021jlq}
Juan Garc\'\i{}a-Bellido, Santiago Jaraba, and Sachiko Kuroyanagi.
\newblock {The stochastic gravitational wave background from close hyperbolic
  encounters of primordial black holes in dense clusters}.
\newblock \emph{Phys. Dark Univ.}, 36:\penalty0 101009, 2022.
\newblock \doi{10.1016/j.dark.2022.101009}.

\bibitem[Garcia-Bellido and Nesseris(2017)]{Garcia-Bellido:2017qal}
Juan Garcia-Bellido and Savvas Nesseris.
\newblock {Gravitational wave bursts from Primordial Black Hole hyperbolic
  encounters}.
\newblock \emph{Phys. Dark Univ.}, 18:\penalty0 123--126, 2017.
\newblock \doi{10.1016/j.dark.2017.10.002}.

\bibitem[Garc{\'\i}a-Bellido and Nesseris(2018)]{Garcia-Bellido:2017knh}
Juan Garc{\'\i}a-Bellido and Savvas Nesseris.
\newblock {Gravitational wave energy emission and detection rates of Primordial
  Black Hole hyperbolic encounters}.
\newblock \emph{Phys. Dark Univ.}, 21:\penalty0 61--69, 2018.
\newblock \doi{10.1016/j.dark.2018.06.001}.

\bibitem[Hansen(1972)]{Hansen1972}
R~O Hansen.
\newblock Post-newtonian gravitational radiation from point masses in a
  hyperbolic kepler orbit*.
\newblock \emph{PHYSICAL REVIEW D VOLUME}, 5, 1972.

\bibitem[{Turner}(1977)]{1977ApJ...216..610T}
M.~{Turner}.
\newblock {Gravitational radiation from point-masses in unbound orbits:
  Newtonian results.}
\newblock \emph{\apj}, 216:\penalty0 610--619, September 1977.
\newblock \doi{10.1086/155501}.

\bibitem[Cho et~al.(2018)Cho, Gopakumar, Haney, and Lee]{Cho:2018upo}
Gihyuk Cho, Achamveedu Gopakumar, Maria Haney, and Hyung~Mok Lee.
\newblock {Gravitational waves from compact binaries in post-Newtonian accurate
  hyperbolic orbits}.
\newblock \emph{Phys. Rev. D}, 98\penalty0 (2):\penalty0 024039, 2018.
\newblock \doi{10.1103/PhysRevD.98.024039}.

\bibitem[Cho et~al.(2022)Cho, Porto, and Yang]{Cho:2022syn}
Gihyuk Cho, Rafael~A. Porto, and Zixin Yang.
\newblock {Gravitational radiation from inspiralling compact objects: Spin
  effects to the fourth post-Newtonian order}.
\newblock \emph{Phys. Rev. D}, 106\penalty0 (10):\penalty0 L101501, 2022.
\newblock \doi{10.1103/PhysRevD.106.L101501}.

\bibitem[{Thorne} and {Kovacs}(1975)]{1975ApJ...200..245T}
K.~S. {Thorne} and S.~J. {Kovacs}.
\newblock {The generation of gravitational waves. I. Weak-field sources.}
\newblock \emph{\apj}, 200:\penalty0 245--262, September 1975.
\newblock \doi{10.1086/153783}.

\bibitem[{Crowley} and {Thorne}(1977)]{1977ApJ...215..624C}
R.~J. {Crowley} and K.~S. {Thorne}.
\newblock {The generation of gravitational waves. II. The postlinear formation
  revisited.}
\newblock \emph{\apj}, 215:\penalty0 624--635, July 1977.
\newblock \doi{10.1086/155397}.

\bibitem[{Kovacs} and {Thorne}(1977)]{1977ApJ...217..252K}
S.~J. {Kovacs} and K.~S. {Thorne}.
\newblock {The generation of gravitational waves. III. Derivation of
  bremsstrahlung formulae.}
\newblock \emph{\apj}, 217:\penalty0 252--280, October 1977.
\newblock \doi{10.1086/155576}.

\bibitem[{Kovacs} and {Thorne}(1978)]{1978ApJ...224...62K}
Jr. {Kovacs}, S.~J. and K.~S. {Thorne}.
\newblock {The generation of gravitational waves. IV. Bremsstrahlung.}
\newblock \emph{\apj}, 224:\penalty0 62--85, August 1978.
\newblock \doi{10.1086/156350}.

\bibitem[Vines(2018)]{Vines:2017hyw}
Justin Vines.
\newblock {Scattering of two spinning black holes in post-Minkowskian gravity,
  to all orders in spin, and effective-one-body mappings}.
\newblock \emph{Class. Quant. Grav.}, 35\penalty0 (8):\penalty0 084002, 2018.
\newblock \doi{10.1088/1361-6382/aaa3a8}.

\bibitem[Bae et~al.(2017)Bae, Lee, Kang, and Hansen]{Bae:2017crk}
Yeong-Bok Bae, Hyung~Mok Lee, Gungwon Kang, and Jakob Hansen.
\newblock {Gravitational radiation driven capture in unequal mass black hole
  encounters}.
\newblock \emph{Phys. Rev. D}, 96\penalty0 (8):\penalty0 084009, 2017.
\newblock \doi{10.1103/PhysRevD.96.084009}.

\bibitem[Bae et~al.(2020)Bae, Lee, and Kang]{Bae:2020hla}
Y.~Bae, H.~Lee, and G.~Kang.
\newblock {Gravitational Wave Capture in Spinning Black Hole Encounters}.
\newblock \emph{\apj}, 900\penalty0 (2):\penalty0 175, 2020.
\newblock \doi{10.3847/1538-4357/aba82b}.

\bibitem[Bae et~al.(2023)Bae, Hyun, and Kang]{Bae:2023sww}
Yeong-Bok Bae, Young-Hwan Hyun, and Gungwon Kang.
\newblock {Ringdown gravitational waves from close scattering of two black
  holes}, 10 2023.
\newblock {Preprint arXiv:2310.18686}.

\bibitem[Favata(2011)]{Favata:2011qi}
Marc Favata.
\newblock {The Gravitational-wave memory from eccentric binaries}.
\newblock \emph{Phys. Rev. D}, 84:\penalty0 124013, 2011.
\newblock \doi{10.1103/PhysRevD.84.124013}.

\bibitem[Thorne(1992)]{PhysRevD.45.520}
K.~S. Thorne.
\newblock Gravitational-wave bursts with memory: The christodoulou effect.
\newblock \emph{Phys. Rev. D}, 45:\penalty0 520--524, Jan 1992.
\newblock \doi{10.1103/PhysRevD.45.520}.
\newblock URL \url{https://link.aps.org/doi/10.1103/PhysRevD.45.520}.

\bibitem[Kocsis et~al.(2006)Kocsis, Gaspar, and Marka]{Kocsis:2006hq}
B.~Kocsis, M.~E. Gaspar, and S.~Marka.
\newblock {Detection rate estimates of gravity-waves emitted during parabolic
  encounters of stellar black holes in globular clusters}.
\newblock \emph{Astrophys. J.}, 648:\penalty0 411--429, 2006.
\newblock \doi{10.1086/505641}.

\bibitem[Barsotti et~al.(2018)Barsotti, McCuller, Evans, and
  Fritschel]{barsotti2018a+}
L.~Barsotti, L.~McCuller, M.~Evans, and P.~Fritschel.
\newblock The a+ design curve.
\newblock \emph{LIGO Document}, 1800042:\penalty0 2018, 2018.

\bibitem[Quinlan and Shapiro(1989)]{Quinlan1989}
Gerald~D Quinlan and Stuart~L Shapiro.
\newblock Dynamical evolution of dense clusters of compact stars.
\newblock \emph{The Astrophysical Journal}, 343:\penalty0 725--749, 1989.

\bibitem[Li et~al.(2022)Li, Lai, and Rodet]{Li2022}
Jiaru Li, Dong Lai, and Laetitia Rodet.
\newblock Long-term evolution of tightly packed stellar black holes in agn
  disks: Formation of merging black hole binaries via close encounters.
\newblock \emph{The Astrophysical Journal}, 934:\penalty0 154, 8 2022.
\newblock ISSN 0004-637X.
\newblock \doi{10.3847/1538-4357/ac7c0d}.

\bibitem[Codazzo et~al.(2023)Codazzo, Di~Giovanni, Harms, Dall'Amico, and
  Mapelli]{Codazzo:2022aqj}
Elena Codazzo, Matteo Di~Giovanni, Jan Harms, Marco Dall'Amico, and Michela
  Mapelli.
\newblock {Study on the detectability of gravitational radiation from
  single-binary encounters between black holes in nuclear star clusters: The
  case of hyperbolic flybys}.
\newblock \emph{Phys. Rev. D}, 107\penalty0 (2):\penalty0 023023, 2023.
\newblock \doi{10.1103/PhysRevD.107.023023}.

\bibitem[Cahillane and Mansell(2022)]{Cahillane:2022pqm}
C.~Cahillane and G.~Mansell.
\newblock {Review of the Advanced LIGO Gravitational Wave Observatories Leading
  to Observing Run Four}.
\newblock \emph{Galaxies}, 10\penalty0 (1):\penalty0 36, 2022.
\newblock \doi{10.3390/galaxies10010036}.

\bibitem[T15(2023)]{T1500293}
{Unofficial sensitivity curves (ASD) for aLIGO, Kagra, Virgo, Voyager, Cosmic
  Explorer and ET}, 2023.
\newblock URL \url{https://dcc.ligo.org/LIGO-T1500293-v13/public}.

\bibitem[Cornish and Littenberg(2015)]{Cornish:2014kda}
Neil~J. Cornish and Tyson~B. Littenberg.
\newblock {BayesWave: Bayesian Inference for Gravitational Wave Bursts and
  Instrument Glitches}.
\newblock \emph{Class. Quant. Grav.}, 32\penalty0 (13):\penalty0 135012, 2015.
\newblock \doi{10.1088/0264-9381/32/13/135012}.

\bibitem[Cornish et~al.(2021)Cornish, Littenberg, B\'ecsy, Chatziioannou,
  Clark, Ghonge, and Millhouse]{Cornish:2020dwh}
Neil~J. Cornish, Tyson~B. Littenberg, Bence B\'ecsy, Katerina Chatziioannou,
  James~A. Clark, Sudarshan Ghonge, and Margaret Millhouse.
\newblock {BayesWave analysis pipeline in the era of gravitational wave
  observations}.
\newblock \emph{Phys. Rev. D}, 103\penalty0 (4):\penalty0 044006, 2021.
\newblock \doi{10.1103/PhysRevD.103.044006}.

\bibitem[Moore et~al.(2015)Moore, Cole, and Berry]{Moore:2014lga}
C.~J. Moore, R.~H. Cole, and C.~P.~L. Berry.
\newblock {Gravitational-wave sensitivity curves}.
\newblock \emph{Class. Quant. Grav.}, 32\penalty0 (1):\penalty0 015014, 2015.
\newblock \doi{10.1088/0264-9381/32/1/015014}.

\bibitem[Berg\'e et~al.(2019)Berg\'e, Massey, Baghi, and
  Touboul]{Berge:2019nyt}
Joel Berg\'e, Richard Massey, Quentin Baghi, and Pierre Touboul.
\newblock {Exponential shapelets: basis functions for data analysis of isolated
  features}.
\newblock \emph{Mon. Not. Roy. Astron. Soc.}, 486\penalty0 (1):\penalty0
  544--559, 2019.
\newblock \doi{10.1093/mnras/stz787}.

\bibitem[{Kanner} et~al.(2016){Kanner}, {Littenberg}, {Cornish}, {Millhouse},
  {Xhakaj}, {Salemi}, {Drago}, {Vedovato}, and {Klimenko}]{2016PhRvD..93b2002K}
Jonah~B. {Kanner}, Tyson~B. {Littenberg}, Neil {Cornish}, Meg {Millhouse}, Enia
  {Xhakaj}, Francesco {Salemi}, Marco {Drago}, Gabriele {Vedovato}, and Sergey
  {Klimenko}.
\newblock {Leveraging waveform complexity for confident detection of
  gravitational waves}.
\newblock \emph{\prd}, 93\penalty0 (2):\penalty0 022002, January 2016.
\newblock \doi{10.1103/PhysRevD.93.022002}.

\bibitem[Drago et~al.(2021)]{Drago:2020kic}
M.~Drago et~al.
\newblock {Coherent WaveBurst, a pipeline for unmodeled gravitational-wave data
  analysis}.
\newblock \emph{SoftwareX}, 14:\penalty0 100678, 2021.
\newblock ISSN 2352-7110.
\newblock \doi{https://doi.org/10.1016/j.softx.2021.100678}.

\bibitem[Ghonge et~al.(2020)Ghonge, Chatziioannou, Clark, Littenberg,
  Millhouse, Cadonati, and Cornish]{Ghonge:2020suv}
Sudarshan Ghonge, Katerina Chatziioannou, James~A. Clark, Tyson Littenberg,
  Margaret Millhouse, Laura Cadonati, and Neil Cornish.
\newblock {Reconstructing gravitational wave signals from binary black hole
  mergers with minimal assumptions}.
\newblock \emph{Phys. Rev. D}, 102\penalty0 (6):\penalty0 064056, 2020.
\newblock \doi{10.1103/PhysRevD.102.064056}.

\bibitem[Millhouse et~al.(2018)Millhouse, Cornish, and
  Littenberg]{Millhouse:2018dgi}
M.~Millhouse, N.~Cornish, and T.~Littenberg.
\newblock {Bayesian reconstruction of gravitational wave bursts using
  chirplets}.
\newblock \emph{Phys. Rev. D}, 97\penalty0 (10):\penalty0 104057, 2018.
\newblock \doi{10.1103/PhysRevD.97.104057}.

\bibitem[Biwer et~al.(2019)Biwer, Capano, De, Cabero, Brown, Nitz, and
  Raymond]{Biwer:2018osg}
C.~M. Biwer, C.~D. Capano, S.~De, M.~Cabero, D.~A. Brown, A.~H. Nitz, and
  V.~Raymond.
\newblock {PyCBC Inference: A Python-based parameter estimation toolkit for
  compact binary coalescence signals}.
\newblock \emph{Publ. Astron. Soc. Pac.}, 131\penalty0 (996):\penalty0 024503,
  2019.
\newblock \doi{10.1088/1538-3873/aaef0b}.

\bibitem[Veitch et~al.(2015)Veitch, Raymond, Farr, Farr, Graff, Vitale, Aylott,
  Blackburn, Christensen, Coughlin, Del~Pozzo, Feroz, Gair, Haster, Kalogera,
  Littenberg, Mandel, O'Shaughnessy, Pitkin, Rodriguez, R\"over, Sidery, Smith,
  Van Der~Sluys, Vecchio, Vousden, and Wade]{PhysRevD.91.042003}
J.~Veitch, V.~Raymond, B.~Farr, W.~Farr, P.~Graff, S.~Vitale, B.~Aylott,
  K.~Blackburn, N.~Christensen, M.~Coughlin, W.~Del~Pozzo, F.~Feroz, J.~Gair,
  C.-J. Haster, V.~Kalogera, T.~Littenberg, I.~Mandel, R.~O'Shaughnessy,
  M.~Pitkin, C.~Rodriguez, C.~R\"over, T.~Sidery, R.~Smith, M.~Van Der~Sluys,
  A.~Vecchio, W.~Vousden, and L.~Wade.
\newblock Parameter estimation for compact binaries with ground-based
  gravitational-wave observations using the lalinference software library.
\newblock \emph{Phys. Rev. D}, 91:\penalty0 042003, Feb 2015.
\newblock \doi{10.1103/PhysRevD.91.042003}.
\newblock URL \url{https://link.aps.org/doi/10.1103/PhysRevD.91.042003}.

\bibitem[{Cannon} et~al.(2021){Cannon}, {Caudill}, {Chan}, {Cousins},
  {Creighton}, {Ewing}, {Fong}, {Godwin}, {Hanna}, {Hooper}, {Huxford},
  {Magee}, {Meacher}, {Messick}, {Morisaki}, {Mukherjee}, {Ohta}, {Pace},
  {Privitera}, {de Ruiter}, {Sachdev}, {Singer}, {Singh}, {Tapia}, {Tsukada},
  {Tsuna}, {Tsutsui}, {Ueno}, {Viets}, {Wade}, and {Wade}]{2021SoftX..1400680C}
K.~{Cannon}, S.~{Caudill}, C.~{Chan}, B.~{Cousins}, J.~D.~E. {Creighton},
  B.~{Ewing}, H.~{Fong}, P.~{Godwin}, C.~{Hanna}, S.~{Hooper}, R.~{Huxford},
  R.~{Magee}, D.~{Meacher}, C.~{Messick}, S.~{Morisaki}, D.~{Mukherjee},
  H.~{Ohta}, A.~{Pace}, S.~{Privitera}, I.~{de Ruiter}, S.~{Sachdev},
  L.~{Singer}, D.~{Singh}, R.~{Tapia}, L.~{Tsukada}, D.~{Tsuna}, T.~{Tsutsui},
  K.~{Ueno}, A.~{Viets}, L.~{Wade}, and M.~{Wade}.
\newblock {GstLAL: A software framework for gravitational wave discovery}.
\newblock \emph{SoftwareX}, 14:\penalty0 100680, June 2021.
\newblock \doi{10.1016/j.softx.2021.100680}.

\bibitem[Abbott et~al.(2019{\natexlab{a}})]{LIGOScientific:2018mvr}
B.~P. Abbott et~al.
\newblock {GWTC-1: A Gravitational-Wave Transient Catalog of Compact Binary
  Mergers Observed by LIGO and Virgo during the First and Second Observing
  Runs}.
\newblock \emph{Phys. Rev. X}, 9\penalty0 (3):\penalty0 031040,
  2019{\natexlab{a}}.
\newblock \doi{10.1103/PhysRevX.9.031040}.

\bibitem[Abbott et~al.(2021{\natexlab{c}})]{LIGOScientific:2020ibl}
R.~Abbott et~al.
\newblock {GWTC-2: Compact Binary Coalescences Observed by LIGO and Virgo
  During the First Half of the Third Observing Run}.
\newblock \emph{Phys. Rev. X}, 11:\penalty0 021053, 2021{\natexlab{c}}.
\newblock \doi{10.1103/PhysRevX.11.021053}.

\bibitem[Abbott et~al.(2021{\natexlab{d}})]{LIGOScientific:2021usb}
R.~Abbott et~al.
\newblock {GWTC-2.1: Deep Extended Catalog of Compact Binary Coalescences
  Observed by LIGO and Virgo During the First Half of the Third Observing Run},
  8 2021{\natexlab{d}}.
\newblock {Preprint arXiv:2108.01045}.

\bibitem[Abbott et~al.(2023)]{KAGRA:2021vkt}
R.~Abbott et~al.
\newblock {GWTC-3: Compact Binary Coalescences Observed by LIGO and Virgo
  during the Second Part of the Third Observing Run}.
\newblock \emph{Phys. Rev. X}, 13\penalty0 (4):\penalty0 041039, 2023.
\newblock \doi{10.1103/PhysRevX.13.041039}.

\bibitem[Abbott et~al.(2019{\natexlab{b}})]{LIGOScientific:2019fpa}
B.~P. Abbott et~al.
\newblock {Tests of General Relativity with the Binary Black Hole Signals from
  the LIGO-Virgo Catalog GWTC-1}.
\newblock \emph{Phys. Rev. D}, 100\penalty0 (10):\penalty0 104036,
  2019{\natexlab{b}}.
\newblock \doi{10.1103/PhysRevD.100.104036}.

\bibitem[Abbott et~al.(2021{\natexlab{e}})]{LIGOScientific:2020tif}
R.~Abbott et~al.
\newblock {Tests of general relativity with binary black holes from the second
  LIGO-Virgo gravitational-wave transient catalog}.
\newblock \emph{Phys. Rev. D}, 103\penalty0 (12):\penalty0 122002,
  2021{\natexlab{e}}.
\newblock \doi{10.1103/PhysRevD.103.122002}.

\bibitem[Pankow et~al.(2018)Pankow, Chatziioannou, Chase, Littenberg, Evans,
  McIver, Cornish, Haster, Kanner, Raymond, Vitale, and
  Zimmerman]{PhysRevD.98.084016}
C.~Pankow, K.~Chatziioannou, E.~A. Chase, T.~B. Littenberg, M.~Evans,
  J.~McIver, N.~J. Cornish, C.~Haster, J.~Kanner, V.~Raymond, S.~Vitale, and
  A.~Zimmerman.
\newblock Mitigation of the instrumental noise transient in gravitational-wave
  data surrounding gw170817.
\newblock \emph{Phys. Rev. D}, 98:\penalty0 084016, Oct 2018.
\newblock \doi{10.1103/PhysRevD.98.084016}.
\newblock URL \url{https://link.aps.org/doi/10.1103/PhysRevD.98.084016}.

\bibitem[Chatziioannou et~al.(2021)Chatziioannou, Cornish, Wijngaarden, and
  Littenberg]{Chatziioannou:2021ezd}
K.~Chatziioannou, N.~Cornish, M.~Wijngaarden, and T.~B. Littenberg.
\newblock {Modeling compact binary signals and instrumental glitches in
  gravitational wave data}.
\newblock \emph{Phys. Rev. D}, 103\penalty0 (4):\penalty0 044013, 2021.
\newblock \doi{10.1103/PhysRevD.103.044013}.

\bibitem[Ghonge et~al.(2023)Ghonge, Brandt, Sullivan, Millhouse, Chatziioannou,
  Clark, Littenberg, Cornish, Hourihane, and Cadonati]{Ghonge:2023ksb}
S.~Ghonge, J.~Brandt, J.~M. Sullivan, M.~Millhouse, K.~Chatziioannou, J.~A.
  Clark, T.~Littenberg, N.~Cornish, S.~Hourihane, and L.~Cadonati.
\newblock {Characterizing the efficacy of methods to subtract terrestrial
  transient noise near gravitational wave events and the effects on parameter
  estimation}, 11 2023.
\newblock {Preprint arXiv:2311.09159}.

\bibitem[Chassande-Mottin et~al.(2010)Chassande-Mottin, Miele, Mohapatra, and
  Cadonati]{Chassande-Mottin:2010hsa}
Eric Chassande-Mottin, Miriam Miele, Satya Mohapatra, and Laura Cadonati.
\newblock {Detection of gravitational-wave bursts with chirplet-like template
  families}.
\newblock \emph{Class. Quant. Grav.}, 27:\penalty0 194017, 2010.
\newblock \doi{10.1088/0264-9381/27/19/194017}.

\bibitem[{B{\'e}csy} et~al.(2017){B{\'e}csy}, {Raffai}, {Cornish}, {Essick},
  {Kanner}, {Katsavounidis}, {Littenberg}, {Millhouse}, and
  {Vitale}]{Becsy2017}
B.~{B{\'e}csy}, P.~{Raffai}, N.~J. {Cornish}, R.~{Essick}, J.~{Kanner},
  E.~{Katsavounidis}, T.~B. {Littenberg}, M.~{Millhouse}, and S.~{Vitale}.
\newblock {Parameter Estimation for Gravitational-wave Bursts with the
  BayesWave Pipeline}.
\newblock \emph{\apj}, 839\penalty0 (1):\penalty0 15, April 2017.
\newblock \doi{10.3847/1538-4357/aa63ef}.

\bibitem[{Pannarale} et~al.(2019){Pannarale}, {Macas}, and
  {Sutton}]{Pannarale2018}
F.~{Pannarale}, R.~{Macas}, and P.~J. {Sutton}.
\newblock {Bayesian inference analysis of unmodelled gravitational-wave
  transients}.
\newblock \emph{Classical and Quantum Gravity}, 36\penalty0 (3):\penalty0
  035011, February 2019.
\newblock \doi{10.1088/1361-6382/aaf76d}.

\bibitem[{B{\'e}csy} et~al.(2020){B{\'e}csy}, {Raffai}, {Gill}, {Littenberg},
  {Millhouse}, and {Szczepa{\'n}czyk}]{Becsy2020}
B.~{B{\'e}csy}, P.~{Raffai}, K.~{Gill}, Tyson~B. {Littenberg}, M.~{Millhouse},
  and M.~J. {Szczepa{\'n}czyk}.
\newblock {Interpreting gravitational-wave burst detections: constraining
  source properties without astrophysical models}.
\newblock \emph{Classical and Quantum Gravity}, 37\penalty0 (10):\penalty0
  105011, May 2020.
\newblock \doi{10.1088/1361-6382/ab7ee2}.

\bibitem[{Henshaw} et~al.(2024){Henshaw}, {Arogeti}, {Heranval}, and
  {Cadonati}]{2024arXiv240216533H}
C.~{Henshaw}, M.~{Arogeti}, A.~{Heranval}, and L.~{Cadonati}.
\newblock {Visualization of frequency structures in gravitational wave
  signals}.
\newblock \emph{arXiv e-prints}, art. arXiv:2402.16533, February 2024.
\newblock \doi{10.48550/arXiv.2402.16533}.

\bibitem[Baghi et~al.(2022)Baghi, Korsakova, Slutsky, Castelli, Karnesis, and
  Bayle]{Baghi:2021tfd}
Quentin Baghi, Natalia Korsakova, Jacob Slutsky, Eleonora Castelli, Nikolaos
  Karnesis, and Jean-Baptiste Bayle.
\newblock {Detection and characterization of instrumental transients in LISA
  Pathfinder and their projection to LISA}.
\newblock \emph{Phys. Rev. D}, 105\penalty0 (4):\penalty0 042002, 2022.
\newblock \doi{10.1103/PhysRevD.105.042002}.

\bibitem[Refregier(2003)]{Refregier:2001fd}
A.~Refregier.
\newblock {Shapelets: I. a method for image analysis}.
\newblock \emph{Mon. Not. Roy. Astron. Soc.}, 338:\penalty0 35, 2003.
\newblock \doi{10.1046/j.1365-8711.2003.05901.x}.

\bibitem[Refregier and Bacon(2003)]{Refregier:2001fe}
A.~Refregier and D.~Bacon.
\newblock {Shapelets. 2. A method for weak lensing measurements}.
\newblock \emph{Mon. Not. Roy. Astron. Soc.}, 338:\penalty0 48, 2003.
\newblock \doi{10.1046/j.1365-8711.2003.05902.x}.

\bibitem[{Birrer} et~al.(2015){Birrer}, {Amara}, and
  {Refregier}]{2015ApJ...813..102B}
Simon {Birrer}, Adam {Amara}, and Alexandre {Refregier}.
\newblock {Gravitational Lens Modeling with Basis Sets}.
\newblock \emph{\apj}, 813\penalty0 (2):\penalty0 102, November 2015.
\newblock \doi{10.1088/0004-637X/813/2/102}.

\bibitem[{Hoekstra} et~al.(2005){Hoekstra}, {Wu}, and
  {Udalski}]{2005ApJ...626.1070H}
Henk {Hoekstra}, Yanqin {Wu}, and Andrzej {Udalski}.
\newblock {An Algorithm to Detect Blends with Eclipsing Binaries in Planet
  Transit Searches}.
\newblock \emph{\apj}, 626\penalty0 (2):\penalty0 1070--1078, June 2005.
\newblock \doi{10.1086/430299}.

\bibitem[{Amara} and {Quanz}(2012)]{2012MNRAS.427..948A}
Adam {Amara} and Sascha~P. {Quanz}.
\newblock {PYNPOINT: an image processing package for finding exoplanets}.
\newblock \emph{\mnras}, 427\penalty0 (2):\penalty0 948--955, December 2012.
\newblock \doi{10.1111/j.1365-2966.2012.21918.x}.

\bibitem[{Ellis} and {Cornish}(2016)]{2016PhRvD..93h4048E}
J.~A. {Ellis} and N.~J. {Cornish}.
\newblock {Transdimensional Bayesian approach to pulsar timing noise analysis}.
\newblock \emph{\prd}, 93\penalty0 (8):\penalty0 084048, April 2016.
\newblock \doi{10.1103/PhysRevD.93.084048}.

\bibitem[{Desvignes} et~al.(2016){Desvignes}, {Caballero}, {Lentati},
  {Verbiest}, {Champion}, {Stappers}, {Janssen}, {Lazarus}, {Os{\l}owski},
  {Babak}, {Bassa}, {Brem}, {Burgay}, {Cognard}, {Gair}, {Graikou},
  {Guillemot}, {Hessels}, {Jessner}, {Jordan}, {Karuppusamy}, {Kramer},
  {Lassus}, {Lazaridis}, {Lee}, {Liu}, {Lyne}, {McKee}, {Mingarelli},
  {Perrodin}, {Petiteau}, {Possenti}, {Purver}, {Rosado}, {Sanidas}, {Sesana},
  {Shaifullah}, {Smits}, {Taylor}, {Theureau}, {Tiburzi}, {van Haasteren}, and
  {Vecchio}]{2016MNRAS.458.3341D}
G.~{Desvignes}, R.~N. {Caballero}, L.~{Lentati}, J.~P.~W. {Verbiest}, D.~J.
  {Champion}, B.~W. {Stappers}, G.~H. {Janssen}, P.~{Lazarus},
  S.~{Os{\l}owski}, S.~{Babak}, C.~G. {Bassa}, P.~{Brem}, M.~{Burgay},
  I.~{Cognard}, J.~R. {Gair}, E.~{Graikou}, L.~{Guillemot}, J.~W.~T. {Hessels},
  A.~{Jessner}, C.~{Jordan}, R.~{Karuppusamy}, M.~{Kramer}, A.~{Lassus},
  K.~{Lazaridis}, K.~J. {Lee}, K.~{Liu}, A.~G. {Lyne}, J.~{McKee}, C.~M.~F.
  {Mingarelli}, D.~{Perrodin}, A.~{Petiteau}, A.~{Possenti}, M.~B. {Purver},
  P.~A. {Rosado}, S.~{Sanidas}, A.~{Sesana}, G.~{Shaifullah}, R.~{Smits}, S.~R.
  {Taylor}, G.~{Theureau}, C.~{Tiburzi}, R.~{van Haasteren}, and A.~{Vecchio}.
\newblock {High-precision timing of 42 millisecond pulsars with the European
  Pulsar Timing Array}.
\newblock \emph{\mnras}, 458\penalty0 (3):\penalty0 3341--3380, May 2016.
\newblock \doi{10.1093/mnras/stw483}.

\bibitem[{Massey} and {Refregier}(2005)]{2005MNRAS.363..197M}
Richard {Massey} and Alexandre {Refregier}.
\newblock {Polar shapelets}.
\newblock \emph{\mnras}, 363\penalty0 (1):\penalty0 197--210, October 2005.
\newblock \doi{10.1111/j.1365-2966.2005.09453.x}.

\bibitem[Abadie et~al.(2012)]{Abadie_2012}
J.~Abadie et~al.
\newblock All-sky search for gravitational-wave bursts in the second joint
  {LIGO}-virgo run.
\newblock \emph{Physical Review D}, 85\penalty0 (12), June 2012.
\newblock ISSN 1550-2368.
\newblock \doi{10.1103/physrevd.85.122007}.
\newblock URL \url{http://dx.doi.org/10.1103/PhysRevD.85.122007}.

\bibitem[Abbott et~al.(2016)]{PhysRevLett.116.061102}
B.~P. Abbott et~al.
\newblock Observation of gravitational waves from a binary black hole merger.
\newblock \emph{Phys. Rev. Lett.}, 116:\penalty0 061102, Feb 2016.
\newblock \doi{10.1103/PhysRevLett.116.061102}.
\newblock URL \url{https://link.aps.org/doi/10.1103/PhysRevLett.116.061102}.

\bibitem[Torrence and Compo(1998)]{torrence1998practical}
C.~Torrence and G.~P. Compo.
\newblock A practical guide to wavelet analysis.
\newblock \emph{Bulletin of the American Meteorological Society}, 79\penalty0
  (1):\penalty0 61--78, 1998.

\bibitem[Mallat(1999)]{mallat1999wavelet}
S.~Mallat.
\newblock \emph{A Wavelet Tour of Signal Processing}.
\newblock Academic Press, 1999.

\bibitem[O'Leary et~al.(2009)O'Leary, Kocsis, and Loeb]{OLeary:2008myb}
R.~M. O'Leary, B.~Kocsis, and A.~Loeb.
\newblock {Gravitational waves from scattering of stellar-mass black holes in
  galactic nuclei}.
\newblock \emph{Mon. Not. Roy. Astron. Soc.}, 395\penalty0 (4):\penalty0
  2127--2146, 2009.
\newblock \doi{10.1111/j.1365-2966.2009.14653.x}.

\end{thebibliography}
